\documentclass[11pt,english,aps,prc,showpacs,letter,longbibliography,preprint]{revtex4-1}
\usepackage{amsmath}
\usepackage[    bookmarks,
                 bookmarksopen = true,
                 bookmarksnumbered = true,
                 linktocpage,
                 colorlinks = true,
                 linkcolor = blue,
                 urlcolor  = blue,
                 citecolor = blue,
                 anchorcolor = green,
                 hyperindex = true,
                 hyperfigures]
        {hyperref}

\usepackage{enumitem}
\usepackage{graphicx}
\usepackage{esint}
\usepackage[utf8]{inputenc}
\usepackage[T1]{fontenc}
\usepackage{float}

\usepackage{fancyhdr}

\usepackage{ulem}
\usepackage{lineno}
\usepackage{listings}

\DeclareFixedFont{\ttb}{T1}{txtt}{bx}{n}{8} % for bold
\DeclareFixedFont{\ttm}{T1}{txtt}{m}{n}{8}  % for normal

\usepackage{color}
\definecolor{deepblue}{rgb}{0,0,0.5}
\definecolor{deepred}{rgb}{0.6,0,0}
\definecolor{deepgreen}{rgb}{0,0.5,0}

\newcommand\pythonstyle{\lstset{
language=Python,
basicstyle=\ttm,
otherkeywords={self},             % Add keywords here
keywordstyle=\ttb\color{deepblue},
emph={MyClass,__init__},          % Custom highlighting
emphstyle=\ttb\color{deepred},    % Custom highlighting style
stringstyle=\color{deepgreen},
frame=tb,                         % Any extra options here
showstringspaces=false            % 
}}

\lstnewenvironment{python}[1][]
{
\pythonstyle
\lstset{#1}
}
{}

\makeatletter
\usepackage{babel}

\makeatother

\begin{document}

\preprint{Submitted to Chinese Physics C}
%%%%% online xyscan:  http://arohatgi.info/WebPlotDigitizer/app/?

\title{Hydrodynamic description of $D$ meson production in high-energy heavy-ion collisions}

\author{Chi Ding$^{1}$, Wei-Yao Ke$^{2,3}$, Long-Gang Pang$^{1}$\footnote{email: lgpang@ccnu.edu.cn} and Xin-Nian Wang$^{1,2,3}$\footnote{email: xnwang@.lbl.gov}\footnote{Current address: Nuclear Science Division, Lawrence Berkeley National Laboratory, Berkeley, CA 94720, USA}}

\address{$^{1}$Key Laboratory of Quark \& Lepton Physics (MOE) and Institute of Particle Physics, Central China Normal University, Wuhan 430079, China}

\address{$^{2}$Physics Department, University of California, Berkeley, CA 94720, USA}

\address{$^{3}$Nuclear Science Division, Lawrence Berkeley National Laboratory, Berkeley, CA 94720, USA}

\begin{abstract}
  The large values and the constituent-quark-number (NCQ) scaling of the elliptic flow of low-$p_T$ $D$ mesons imply that charm quarks, initially produced through hard processes, might be partially thermalized through the strong interactions with the quark-gluon plasma (QGP) in high-energy heavy-ion collisions. 
  To quantify the degree of thermalization of low-$p_T$ charm quarks, we compare the $D^0$ meson spectra and elliptic flow from a hydrodynamic model to the experimental data as well as transport model simulations. 
  We use an effective charm chemical potential at the freeze-out temperature to account for the initial charm quark production from hard processes and assume that they are thermalized in local comoving frame of the medium before freeze-out.   
  $D^0$ mesons are sampled statistically from the freeze-out hyper-surface of the expanding QGP as described by the event-by-event (3+1)D viscous hydrodynamic model CLVisc. 
  Both hydrodynamic and transport model can describe the elliptic flow of $D^0$ mesons at $p_T<3$ GeV/$c$ as measured in Au+Au collisions at $\sqrt{s_{NN}}=200$ GeV.  
  Though the experimental data on $D^0$ spectra are consistent with the hydrodynamic result at small $p_T\sim 1$ GeV/$c$, they deviate from the hydrodynamic model at high transverse momentum $p_T>2$ GeV/$c$.   
  The diffusion and parton energy loss mechanisms in the transport model can describe the measured spectra reasonably well within the theoretical uncertainty.
  Our comparative study indicates that charm quarks only approach to local thermal equilibrium at small $p_T$ even though they acquire sizable elliptic flow that is comparable to light-quark hadrons at both small and intermediate $p_T$.
\end{abstract}



\maketitle

\section{Introduction}

The large collective flow and momentum anisotropy as manifested in the final hadron spectra in high-energy heavy-ion collisions indicate a strong collectivity of the produced dense and hot nuclear matter during its dynamic evolution~\cite{Gelman:2006xw,Heinz:2008tv}. 
The observed approximate constituent quark number (NCQ) scaling of the elliptic flow for light quark hadrons \cite{Adler:2003kt,Adams:2003am,Abelev:2014pua,Adler:2002tq} and strong jet quenching \cite{Adcox:2001jp,Adler:2003qi,Adler:2002xw,CMS:2012aa} in high-energy heavy-ion collisions at both the Relativistic Heavy-ion Collider (RHIC) and the Large Hadron Collider (LHC) suggest the formation of a hot, deconfined and opaque quark-gluon plasma (QGP). 
Similar phenomena are also observed for heavy quark mesons \cite{Adamczyk:2014uip,Adam:2018inb,Adamczyk:2017xur,ALICE:2012ab,Abelev:2013lca,Sirunyan:2017xss,Sirunyan:2017plt}. 
Experimental data from RHIC and LHC show a large elliptic flow of $D^0$ mesons at low-$p_T$, which also approximately obeys the NCQ scaling as light quark hadrons \cite{Adamczyk:2017xur,Sirunyan:2018toe}. 
Heavy quarks are predominantly produced through initial hard processes and have mass scales much larger than the typical temperature of the QGP medium. 
Therefore, it still remains an interesting question as to whether and to what degree charms quarks become thermalized \cite{Heinz:2008tv} and flow with the QGP due to their strong interaction with the hot medium.

Past studies usually describe the thermalization processes of heavy quarks with transport equations.
Their low-momentum ($p\lesssim M$) dynamics in the QGP medium is usually treated as a Brownian motion \cite{RevModPhys.17.323,Rapp:2009my,Moore:2004tg}, considering masses of heavy quarks being much larger than the typical temperatures in the medium $M\gg T$.
For the transport of heavy quarks with intermediate momentum $p\gtrsim M$, both the collisional and radiative energy loss have to be considered.
For high-momenta heavy quarks, most transport models \cite{Cao:2016gvr,Nahrgang:2016lst,WeiYao2018} consider the gluon radiations and interactions between heavy quarks and the medium as perturbative processes or use transport coefficients with weakly-coupled assumptions as inputs.

It was found that transport models need a large heavy-quark momentum-diffusion parameter ($\hat{q}=0.7...2.7$ GeV${}^2/$fm at $T=300$ MeV and $p=30$ GeV/$c$ \cite{Cao:2018ews}) to describe the observed medium modification of the heavy-quark meson spectra and elliptic flow coefficients \cite{Rapp:2018qla,Cao:2018ews}.
There have been several works trying to evaluate the heavy quark diffusion coefficient at zero momentum non-perturbatively from lattice QCD \cite{Banerjee:2011ra,Ding:2012sp,Francis:2015daa}. 
The estimated interaction strength is comparable to the values from phenomenological determination \cite{Xu:2017obm,WeiYao2018}, which is much larger than the natural expectation of the weakly-coupled theory at the leading order.
This poses questions on the weak-coupling assumption on the nature of the interaction between heavy quarks and the QGP medium.
It is possible that, with a strong coupling, the dynamics of the low-$p_T$ charm quarks in the QGP is better described by a hydrodynamic approach.

In this paper, we will compute the production of low-$p_T$ $D$ mesons in an extreme limit of hydrodynamic evolution with the following assumptions.
First, the number of (anti)charm quarks is conserved during the lifetime of the fireball, i.e, thermal production of heavy quark pairs is negligible. 
Second, the coupling is so strong that the initially produced charm quarks quickly diffuse into the QGP and reach local kinetic equilibrium.
Finally, the differential yield of $D$ mesons is computed using the Cooper-Frye formula \cite{Cooper:1974mv} on the hydrodynamic freeze-out hypersurface, with an effective charm chemical potential to guarantee the charm yield is the same as measured in the experiment.
We compare the resultant $p_T$-dependent spectra and elliptic flow of low-$p_T$ $D$ mesons, which are considered as the manifestation of the extreme limit of complete thermalization of heavy quarks, to the experimental data as well as transport calculations to quantify the degree of heavy-quark thermalization in heavy-ion collisions. 
We will use the (3+1)D hydrodynamic model (CLVisc)\cite{Pang:2018zzo} and a linearized Boltzmann-Langevin model \cite{WeiYao2018} for heavy quark transport. 

The paper is organized as follows. In Section \ref{sec:method}, we introduce the relativistic hydrodynamic model and the Boltzmann-Langevin transport model for the calculations of $p_T$ spectra and elliptic flow of charmed mesons.  
In Section \ref{sec:results}, we first discuss the calibration of the hydro model to the experimental data on light quark hadron spectra and elliptic flow in Au+Au collisions at the RHIC energy. 
We then compare the hydro results on the $p_T$ spectra and elliptic flow of charmed mesons at low $p_T$ to the experimental data and transport calculations and discuss the implication on the degree of thermalization for heavy quarks in the QGP formed in high-energy heavy-ion collisions.

\section{Model Descriptions}
\label{sec:method}
\subsection{Charm meson production in the limit of local kinetic equilibrium}
In the usual paradigm for charmed meson production in high-energy heavy-ion collisions, charm quarks are produced through the initial hard scatterings \cite{Collins:1985gm}. 
These heavy quarks will experience energy loss and momentum diffusion in the QGP through both elastic and inelastic collisions with the medium, which can be modeled as the drag and diffusion coefficients in the Boltzmann-Langevin equations. 
The final $D$ mesons are formed through charm quark fragmentation at high $p_T$ \cite{Braaten:1994bz,Cacciari:2003zu,Cacciari:2012ny} or charm-light quark recombination at low and intermediate $p_T$ \cite{Rapp:2003wn}. 
Such transport models for initial heavy quark and final meson production can describe the charm meson spectra and elliptic flow well in high-energy heavy-ion collisions \cite{Rapp:2018qla,Cao:2018ews}. 
However, it is still interesting to investigate whether and to what degree the heavy quarks achieve kinematic equilibrium.
If these heavy quarks indeed reach local thermal equilibrium, one should expect that the hydrodynamic model can also describe heavy meson spectra and elliptic flow.
Furthermore, a hydrodynamic picture also overcomes the difficulty of the transport equation when the coupling becomes large between charm quarks or mesons and the medium.

In this study, we will consider the extreme scenario and investigate the charmed meson spectra and elliptic flow in the limit of the fully thermalized low-$p_T$ charm quarks.
These low-$p_T$ charm quarks are still produced initially through hard processes, but the interactions are assumed to be strong enough that they quickly diffuse into the medium and lose the memory of their initial distribution in phase space, both the spatial location and momentum of the initial production through hard processes.
In the momentum space, they are assumed to reach full kinetic equilibrium and  comove with the medium, flowing with the strongly coupled quark-gluon plasma as light quarks and gluons. 
Finally, the charm quarks transit to charm hadrons with a phase transition of the bulk medium where we assume the interaction in the hadronic phase is still strong enough to maintain the kinetic equilibrium and the equilibrium ratios of different species of charm mesons.
In this situation, the low-$p_T$ D mesons are produced in the same way as other light hadrons on the freeze-out hypersurface in a relativistic hydrodynamic model.  
The temperature and the fluid velocity profiles on the freeze-out hypersurface are crucial for a reasonable estimate of the $D$-meson spectra and elliptic flow in the limit of a complete heavy-quark thermalization.
We will use the rapidity distribution, $p_T$ spectra, and elliptic flow of charged pions to calibrate the hydrodynamic model. 

For completeness, we have to consider $D^0$ mesons from the feed-down of $D^*$. 
Using the feed-down tables from \cite{Rapp:2003wn,Patrignani:2016xqp},

\begin{eqnarray}
D^*(2007)^0 & \xrightarrow{64.7\%} & D^0(1865) + \pi^0, \\
D^*(2007)^0 & \xrightarrow{35.3\%} & D^0(1865) + \gamma, \\
D^*(2010)^+ & \xrightarrow{68\%} & D^0(1865) + \pi^+
\end{eqnarray}
Notice that the $D^*$'s are vector mesons with spin 1. So the spin degeneracy ratio for $D^* / D$ is  $3:1$.

\subsection{Relativistic hydrodynamics}
The hydrodynamic model we use, CLVisc,  is a (3+1)D viscous hydrodynamic model parallelized on GPU using OpenCL \cite{Pang:2018zzo}. 
The program is well tested against several analytical solutions and can describe the bulk hadron spectra and anisotropic flow in high-energy heavy-ion collisions at top RHIC and LHC energy. 
It simulates the fluid dynamic evolution of the strongly interacting QCD matter created in high-energy heavy-ion collisions by solving the fluid dynamic equations,
\begin{equation}
    \nabla_\mu T^{\mu\nu} = 0, {\ \rm with\ } T^{\mu\nu}= \varepsilon u^\mu u^\nu - P\Delta^{\mu\nu}+\pi^{\mu\nu},
\end{equation}
where $\varepsilon$ is the energy density, $P$ is the pressure as a function of energy density given by the equation of state (EoS), $\Delta^{\mu\nu}=g^{\mu\nu}-u^\mu u^\nu$ is a projection operator, $u^\mu$ is the fluid four-velocity obeying $u^\mu u_\mu=1$ and $\pi^{\mu\nu}$ is the shear stress tensor. 

The initial condition for entropy density distribution in the transverse plane is provided by Trento Monte Carlo model \cite{Moreland:2014oya,Bernhard:2016tnd}. 
An envelope function is used to approximate the longitudinal distribution along the space-time rapidity,
\begin{equation}
    H\left(\eta_{s}\right)=\exp \left[-\frac{\left(|\eta_{s}|-\eta_{w}\right)^{2}}{2 \sigma_{\eta}^{2}} \theta\left(|\eta_{s}|-\eta_{w}\right)\right],
\end{equation}
where $\sigma_{\eta}=1.5$ and $\sigma_{w}=1.3$ are used for Au+Au collisions at $\sqrt{s_{NN}}=200$ GeV.

We have assumed an initial time for the hydrodynamics $\tau_0=0.6$ fm.
In the present study, we use the partial chemical equilibrium EoS with chemical freeze-out temperature 165 MeV and a smooth crossover between a QGP at high temperature and the hadron resonance gas (HRG) EoS at low temperature \cite{Huovinen:2009yb} as inspired by the lattice QCD study.

Baryons and mesons passing through the freeze-out hyper-surface are assumed to obey Fermi-Dirac and Bose-Einstein distributions, respectively. Their momentum distributions are given by the Cooper-Frye formula \cite{Cooper:1974mv},
\begin{equation}
\frac{dN_i}{dy p_Tdp_Td\phi}=\frac{g_i}{(2\pi)^3}\int{p^\mu d\Sigma_\mu f(p\cdot u)}(1+\delta f),
\end{equation}
where $g_i = 2 {\rm spin} + 1$ is the spin degeneracy, $p^{\mu}$ is the four-momenta of produced particles in lab frame, $\Sigma_{\mu}$ is the freeze-out hyper-surface, $f(p\cdot u)$ is the Fermi-Dirac/Bose-Einstein distribution function,
\begin{equation}
f(p\cdot u) =\frac{1}{\exp \left[\left(p \cdot u-\mu_{i}\right) / T_{\mathrm{frz}}\right] \pm 1} ,
\end{equation}
and $\delta f$ is the non-equilibrium correction,
\begin{equation}
\delta f =\left(1 \mp f_{\mathrm{eq}}\right) \frac{p_{\mu} p_{\nu} \pi^{\mu \nu}}{2 T_{\mathrm{frz}}^{2}(\varepsilon+P)}.
\end{equation}
We have chosen the freeze-out temperature $T_{fz}=137MeV$ for light flavor hadrons. The freeze-out temperature for $D$ mesons will be different and we will consider several values to provide an estimate of the uncertainties.

The elliptic flow of $D$ mesons is defined as the second coefficient of the Fourier decomposition of their azimuthal angle distributions
with respect to the event plane of light hadrons,
\begin{equation}
\frac{d^{3} N}{p_{\mathrm{T}} d p_{\mathrm{T}} d y d \phi}=\frac{d^{2} N}{2 \pi p_{\mathrm{T}} d p_{\mathrm{T}} d y}\left[1+\sum_{n=1}^{\infty} 2 v_{n} \cos \left(n\left(\phi-\Psi_{\mathrm{EP}}\right)\right)\right].
\end{equation}

\subsection{Transport approach of heavy flavor evolution}

We will compare the charm meson spectra and flow calculated from the hydrodynamic freeze-out model to that from a transport model~\cite{WeiYao2018}.
The transport approach assumes that heavy quarks, including those with low momentum in the comoving frame of the medium, remain good quasi-particles in the QGP.
Therefore, the dynamics of low-$p_T$ heavy flavors can be described by a Boltzmann-type transport equation,
\begin{eqnarray}
\left(\frac{\partial}{\partial t}+\mathbf{v}\cdot \nabla\right) f_Q(t,\mathbf{x}, \mathbf{p}) = \int \left[\frac{dR}{d\mathbf{q}^3}(p+q,q) f_Q(p+q) - \frac{dR}{d\mathbf{q}^3}(p,q)f_Q(p)\right] d\mathbf{q}^3.
\end{eqnarray}
Here, $f_Q(t, {\bf x}, {\bf p})$ is the phase-space density of heavy flavors: heavy quarks or heavy mesons (heavy baryon is omitted in this study).
$dR(p,q)/dq^3$ is the differential rate for a heavy flavor particle with momentum $p$ to transfer three-momentum $\mathbf{q}$ to the local medium with flow velocity $u$ and temperature $T$.

The hydrodynamic and the transport model have both overlapping and distinct regimes of application.
Both models contain the equilibrium situation. 
Compared to the hydrodynamic approach, a transport model also governs the far-from-equilibrium dynamics of heavy flavor particles.
The traces of off-equilibrium effects can be important in a finite and expanding plasma with moderate coupling between heavy quarks and the medium. In the meantime, hydrodynamics can describe the evolution with large couplings and in the non-quasiparticle regime, which is beyond the applicability of the transport approach.
Therefore, both models provide complimentary pictures to understand the experimental data.

The initial charm quark spectrum that initializes the transport equation is obtained from the perturbative QCD based FONLL \cite{Cacciari:1998it,Cacciari:2001td} calculation with EPPS16 nuclear parton distribution function \cite{Eskola:2016oht}.
In the medium above the pseudo-critical temperature ($T>T_c$), we assume the heavy flavors exist as deconfined heavy quarks. 
The interaction rate consists of two parts with a comparable contribution to heavy quark energy loss at intermediate $p_T$ region: collisional processes and medium-induced gluon radiations.

\paragraph{Collisional process in the QGP} collisional processes are mainly modeled by two-body scatterings between the heavy quark and medium partons,
\begin{eqnarray}
\frac{dR_{i=q,g}}{d\mathbf{q}^3} = \int \overline{|M_{Qi\rightarrow Q'i'}^2|}\frac{d_i f(p_i)}{16\pi^2 E_QE_Q'} \frac{d^3p_i}{2E_i} \frac{d^3p_i'}{2E_i'} \delta^{(4)}(p_Q+p_i-p_Q'-p_i'),
\label{eq:R22}
\end{eqnarray}
where $Q$ ($i$) and $Q'$ ($i'$) labels the initial and final state heavy quark (medium quark or gluon).
$\mathbf{p_Q'}=\mathbf{p_Q}+\mathbf{q}$ is the three-momentum transfer to the heavy quark.
$\overline{|M_{Qi\rightarrow Q'i'}^2|}$ is the squared amplitude of the two-body collision at leading order, averaging over initial-stage quantum number and summed over final-state quantum number.
The $\hat{t}$-channel divergence is screened by using the QCD non-perturbative scale $\Lambda=0.2$ GeV and the Debye screening mass $m_D=\sqrt{6\pi\alpha_s} T$ of a three-flavor plasma,
\begin{eqnarray}
\overline{|M_{Qi\rightarrow Q'i'}^2|} \sim \frac{1}{\hat{t}^2} \rightarrow \frac{1}{(\hat{t}-\Lambda^2)(\hat{t}-m_D^2)}.
\end{eqnarray}
Finally, $f(p_i) = e^{-p_i\cdot u/T}$ is the classical thermal distribution function of the medium parton in plasma with local four-velocity $u$ and temperature $T$, obtained from hydrodynamic simulations. The degeneracy factor for medium quarks and gluons are $d_g = 2 d_A = 16$, and $d_q = 4 N_f N_c =36$.
Besides perturbative scatterings, an additional effective heavy quark diffusion constant of the form
\begin{eqnarray}
\frac{\Delta \kappa_D}{T^3} = A+\frac{B}{ET}
\end{eqnarray}
is introduced to mimic possible non-perturbative contribution whose contribution peaks at low energy and low temperature.
The effective heavy-quark-medium coupling parameter and the parameters in the non-perturbative diffusion constant were tuned in \cite{WeiYao2018} to the suppression and momentum anisotropy of the production of open heavy flavor particles.

\paragraph{Medium-induced radiative process in the QGP phase} 
Heavy quark can radiates additional gluon under the collision with medium partons. This is treated similarly to the rate in Eq. \ref{eq:R22} for collisional processes, using two-to-three-body matrix-elements.
One complication is that energetic gluon radiation in the rest frame of the medium is suppressed due to the QCD Landau-Pomeranchuk-Migdal (LPM) effect. 
In \cite{WeiYao2018}, this was included by restricting the phase-space of the radiated gluon using an interference factor motivated by the Higher-Twist approach $2[1-\cos( t/\tau_f)]$, where $t$ is the time since the last gluon radiation, while $\tau_f=2x(1-x)E_Q/(k_\perp^2+x^2M^2)$ is the gluon formation time.
Despite the LPM suppression, the radiative energy loss was found to be equally important in the intermediate $p_T$ region.

\paragraph{Heavy quark hadronization and hadronic rescattering} at the pseudo-critical temperature, charm quarks hadronize to $D$-mesons through both fragmentation and recombination mechanisms \cite{Cao:2013ita}.
In the hadronic phase, $D$-mesons continued to interact with the light mesons via $D$-$\pi$ and $D$-$\rho$ scatterings \cite{Lin:2000jp} as implemented in the Ultra-relativistic Quantum Molecular Dynamics (UrQMD) \cite{Bass:1998ca,Bleicher:1999xi}.

\section{Results}
\label{sec:results}

%%%%%%%%%%%%%%%%%%

\subsection{Model calibration with experimental data}

\begin{figure}[!htp]
\begin{center}
\includegraphics[scale=0.55]{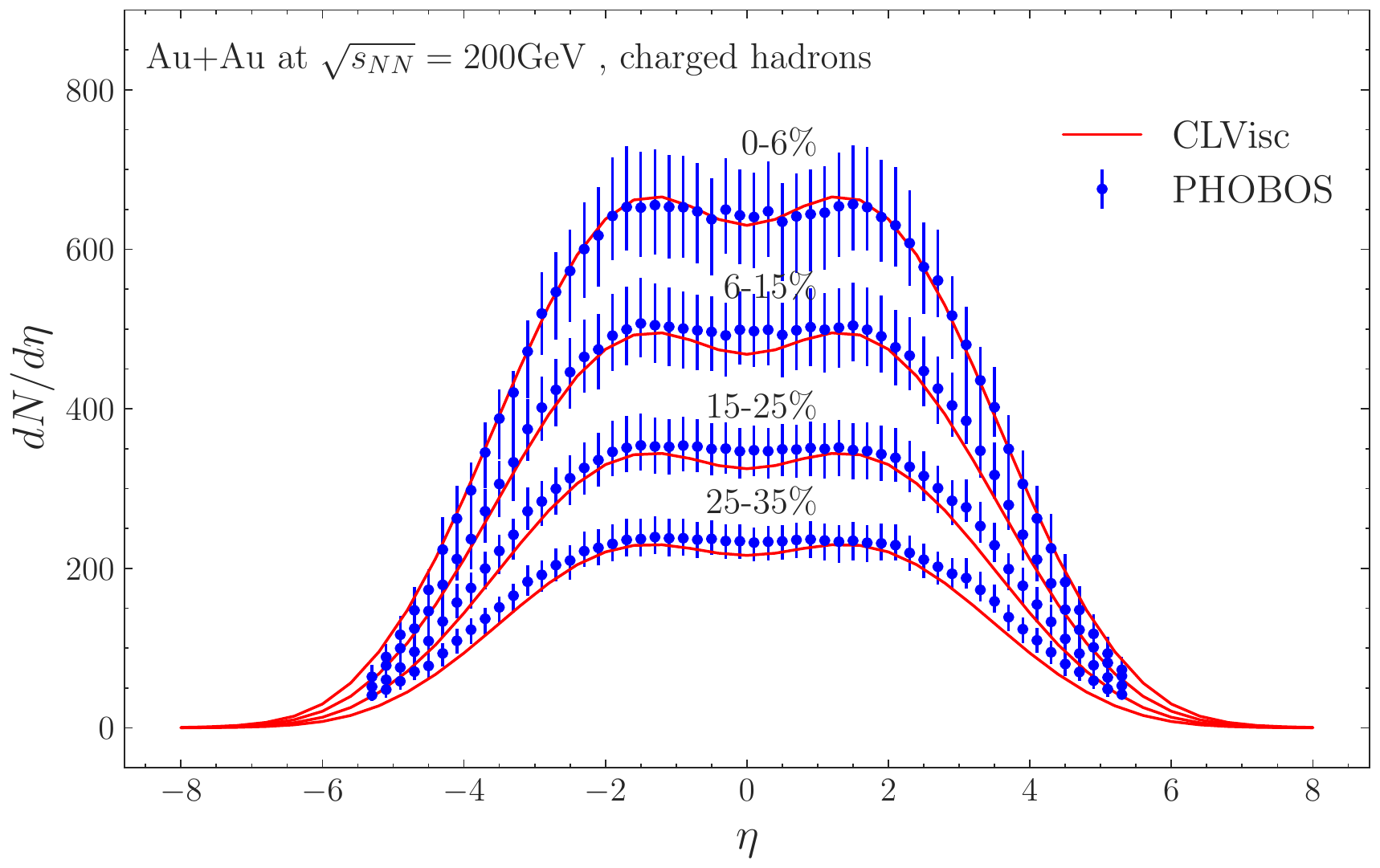}
\end{center}
\caption{The pseudorapidity density of charged hadrons from integrating the Cooper-Frye formula over the freeze-out hyperspace obtained in the CLVisc calculation.
It agrees with data of the PHOBOS experiment for Au + Au collisions  at $\sqrt{s_{NN}}=200$GeV~\cite{Back:2002wb}.}
\label{fig:calib1}
\end{figure}

\begin{figure}[!htp]
    \centering
    \includegraphics[width=0.49\textwidth]{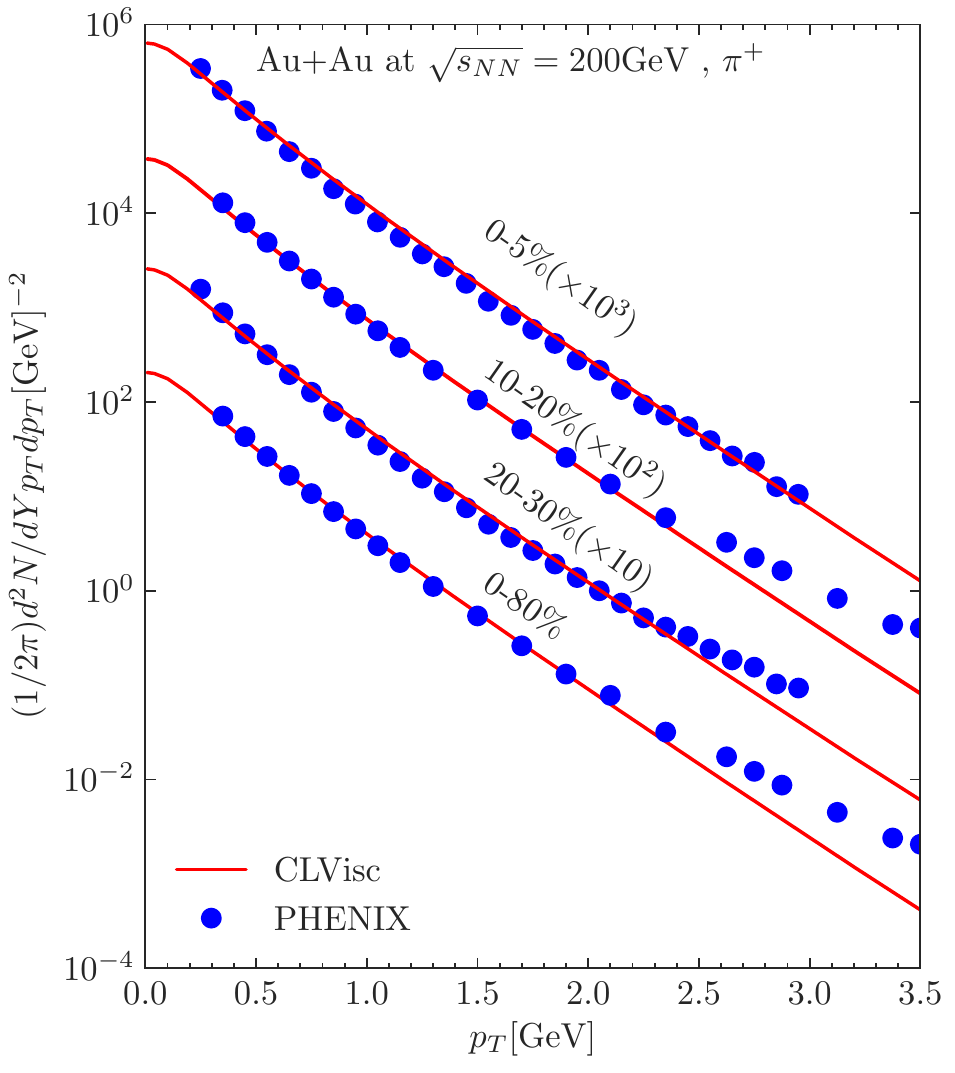}
    \includegraphics[width=0.49\textwidth]{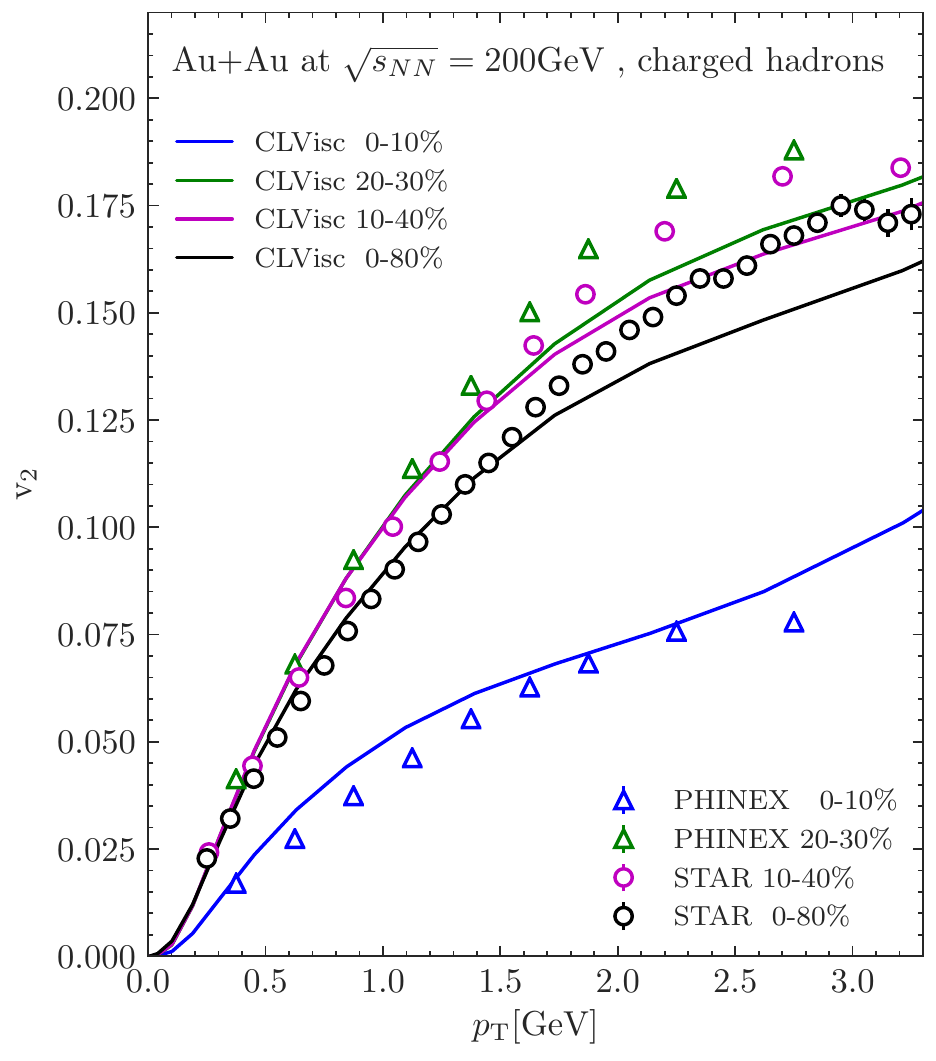} 
    \caption{Left: the transverse momentum spectra of $\pi^+$ in  Au + Au collisions.
    Right: the anisotropic flow coefficient $v_2$ in  Au + Au collisions using the event-plane method.
    In both panels, CLVisc calculations (lines) are compared to data from the PHENIX experiment~\cite{Adler:2003cb} and STAR experiment\cite{Abelev:2008ae}. }
    \label{fig:calib2}
\end{figure}

To calibrate the relativistic hydrodynamic model, we have computed the pseudorapidity density of charged particles $dN_{\textrm{ch}}/d\eta$ as shown in Fig.~\ref{fig:calib1}, the $p_T$ spectra and the elliptic flow $v_2$ of $\pi^+$ as shown in Fig.~\ref{fig:calib2} for central Au+Au collisions with centrality range $0-5\%$ at $\sqrt{s_{NN}}=200$ GeV. 
The calibrated parameters are the scale factor of 53 multiplied to the initial entropy density from the Trento model \cite{Moreland:2014oya,Bernhard:2016tnd}, the starting time for hydro $\tau_0 = 0.6$ fm, the shear viscosity over entropy density ratio $\eta/s = 0.15$, and the freeze-out temperature for light hadrons $T_{fz} = 137$ MeV. These values have been tuned such that the predicted $dN_{\textrm{ch}}/d\eta$, $p_T$ spectra and $v_2(p_T)$ of charged pions from the relativistic hydrodynamics agree with the experimental data.
Relativistic hydrodynamics with the same set of parameters also describes data well at other centralities and we refer readers to Ref. \cite{Pang:2018zzo} for a more detailed description of CLVisc and the parameters. 
These parameters are fixed in the following calculations of the heavy quark meson spectra, except the freeze-out temperatures for $D$ mesons. 
We will vary the value of the $D$ meson freeze-out temperature and study its effect on the final $D$ meson spectra and the elliptic flow.

%%%%%%%%%%%%%%%%
\subsection{Effects of the freeze-out temperature and resonance decay on $D$ spectra}
\begin{figure}[!htp]
    \centering
    \includegraphics[width=0.49\textwidth]{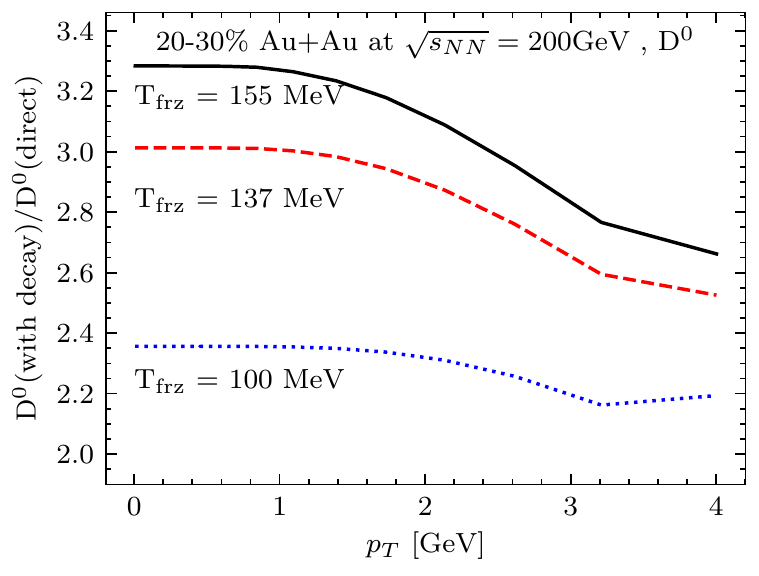} \includegraphics[width=0.49\textwidth]{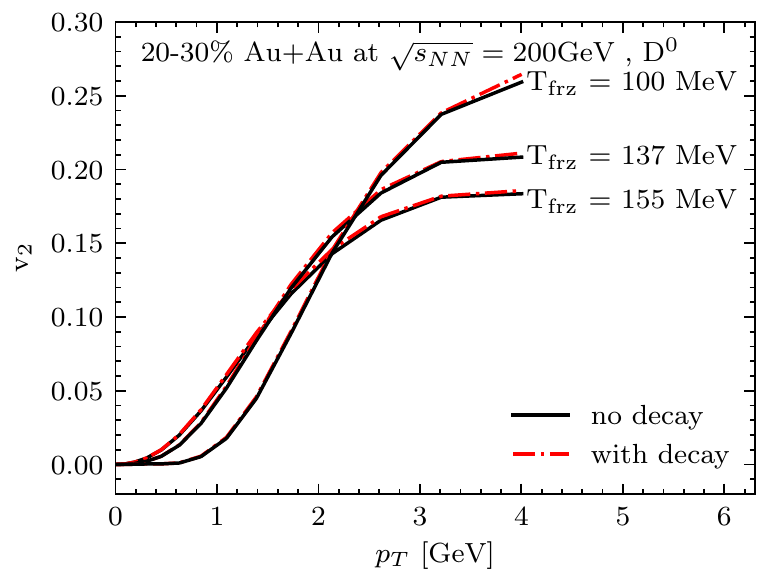}
    \caption{ Left: the freeze-out temperature dependence of the effect of resonance decay on the $D^0$ $p_T$ spectra.
    Right: the freeze-out temperature dependence of the effect of resonance decay on the $D^0$ flow coefficient $v_2$.
    Calculations in both panels use one-shot hydrodynamic simulation of 20-30\% central Au+Au collisions at $\sqrt{s_{NN}}=200$ GeV.}
    \label{fig:reso_contrib}
\end{figure}

The effects of freeze-out temperature and resonance decays on $D^0$ meson production are studied in this section.
In the left panel of Fig.~\ref{fig:reso_contrib}, we show the ratio between $D^0$ $p_T$ spectra with and without contributions from the resonance decays.
In the right panel of Fig.~\ref{fig:reso_contrib}, we compare the $D^0$ meson elliptic flow as a function of $p_T$ with (dot-dashed) and without (solid) resonance decays for three different values of the $D$ meson freeze-out temperatures,  $T_{fz}=$100, 137 and 155 MeV. 
We see that the resonance decay has a much larger effect on the $p_T$ spectra of $D^0$ (left panel) than that on the elliptic flow $v_2$ of $D^0$ (right panel). 
For a given $D$ meson freeze-out temperature, resonance decays from $D^{*}$ contribute more $D^0$ mesons at low $p_T$ than that at high $p_T$. 
The ratio between $D^0$ with and without decay decreases as the $p_T$ increase for all three freeze-out temperatures considered here. 
As one increases freeze-out temperature from $100$ MeV to $137$ MeV and $155$ MeV, not only the $D^0$ yield increases, but also the ratio between $D^0$ with and without decay increases, from about  $2.35$ to $3.0$ and $3.3$ at $p_T<1$ GeV. 
The elliptic flow without decay (black solid), on the other hand, almost overlap that with decays (red dash-dotted) for all three different freeze-out temperatures, therefore, it is insensitive to the contributions from the resonance decays. 
The freeze-out temperature, however, has a large effect on the shape and magnitude of the elliptic flow as a function of transverse momentum. 
At small transverse momentum ($p_T$ < 2 GeV), the elliptic flow increases with the freeze-out temperature.
At large transverse momentum ($p_T$ > 2 GeV), the trend goes the opposite way -- the elliptic flow decreases as the freeze-out temperature increases. 

\subsection{Model and data comparisons of the $p_T$ spectra and $v_2$ of $D^0$ mesons}

\begin{figure}[!htp]
    \centering
    \includegraphics[width=0.49\textwidth]{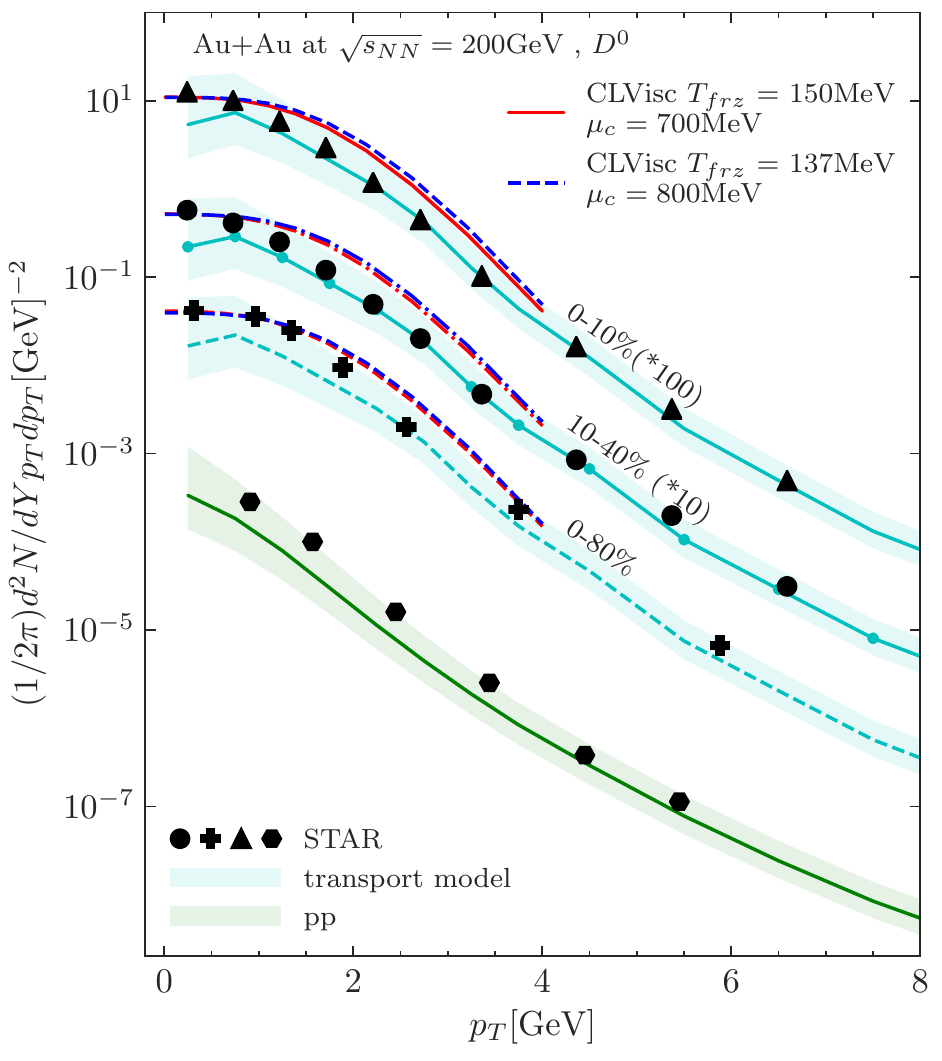}
    \includegraphics[width=0.49\textwidth]{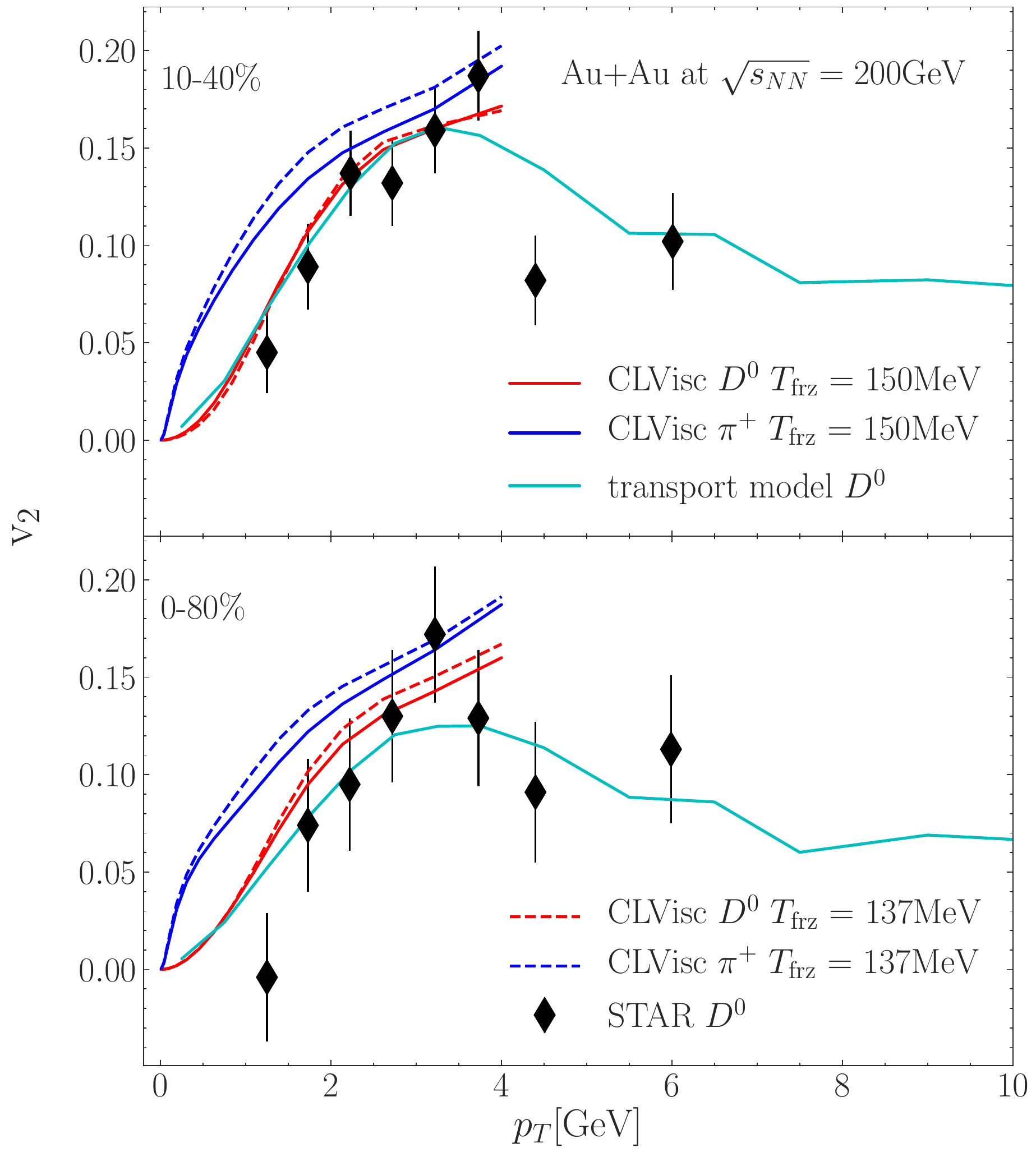} 
    \caption{
    Left: hydrodynamic calculations (lines) of $D^0$ $p_T$ spectra in Au + Au collisions at $\sqrt{s_{\rm NN}}=200$~GeV for three different centrality bins, using two different sets of freeze-out temperatures and effective charm chemical potentials.
    They are compared to transport calculations in the same collision system (blue bands), and data from the STAR experiment~\cite{Adam:2018inb}.
    We also compare the D-meson production in p + p collisions from FONLL calcualtion (green band) \cite{Cacciari:1998it,Cacciari:2001td} to the STAR measurement \cite{Adamczyk:2012af}.
    Right: hydrodynamic calculations of $D^0$ meson $v_2$ are compared to transport calculations and STAR data \cite{Adamczyk:2017xur}.
    }
    \label{fig:v2 and pt spectrum}
\end{figure}

In Fig.~\ref{fig:v2 and pt spectrum}, we show the hydrodynamic calculations of the transverse momentum ($p_T$) spectra (left) and the $p_T$-dependent elliptic flow (right) of $D^0$ mesons in Au + Au collisions at $\sqrt{s_{\rm NN}}=200$ GeV for different centralities as compared with the results of the transport approach and experimental data from the STAR experiment \cite{Adam:2018inb}. 
Since the thermal production of charm quark pairs is negligible in the QGP at the temperatures reached in heavy-ion collisions at the RHIC and LHC colliding energies, we assume the charm quark number from the initial hard processes is conserved via an introduction of charm chemical potential at the freeze-out. 
The value of the chemical potential $\mu_c$ at the freeze-out temperature $T_{frz}$ is adjusted to fit the magnitude of the experimental data on the $p_T$ spectra at low $p_T$ in the most central collisions. 
For $T_{frz}=150$ MeV, the effective charm chemical potential $\mu_c=700$ MeV is found to fit the experimental data on $D^0$ spectra (red) at low $p_T$ ($p_T < 1\ GeV$) for 0-10\% (solid), 10-40\% (dot-dashed) and 0-80\% (dashed) centrality.  
With the same freeze-out temperature and the effective charm chemical potential in the most central collisions, the hydrodynamic model can describe the low $p_T$ spectra well at other centralities. 
However, the hydro calculations over predict the $p_T$ spectra at $p_T>1$ GeV/$c$.

Different hadrons can freeze-out at different temperatures in the hydrodynamic model of hadron production. 
For example, it was shown in the UrQMD model studies that protons freeze-out earlier than pions and kaons \cite{vanHecke:1998yu}.
In the hydro calculation of $D^0$ spectra (red in the left panel), we have assumed that $D^0$ mesons freeze out ($T_{frz}=150$ MeV) earlier than the light quark hadrons ($T_{frz}=137$ MeV). 
If we set the freeze-out temperature of $D^0$ as the same as the light quark hadrons at 137 MeV with a re-adjusted charm chemical potential $\mu_c=800$ MeV, the slope of the $p_T$ spectra becomes slightly larger (blue) due to increased radial expansion, further away from experimental data. 
Increasing the freeze-out temperature will slightly decrease the slope of the spectra at high $p_T$. However, the change is too small to describe the data at high $p_T$.

In the left panel of Fig.~\ref{fig:v2 and pt spectrum}, we further compare to the transport calculations (blue bands) of the $D^0$ spectra in Au+Au collisions.
Before discussing the results in Au-Au collisions, it is necessary to present the FONLL calculation of the baseline $D^0$ spectra (green band) in p+p collisions.
The $D^0$ meson spectra in p+p collisions are computed using the FONLL program \cite{Cacciari:1998it,Cacciari:2001td}, whose major uncertainty stems from the variation of the renormalization scale $m_T/2<\mu_R<2m_T$, where $m_T = \sqrt{M^2+p_T^2}$ is the transverse mass of the heavy quark.
From Fig.~\ref{fig:v2 and pt spectrum}, the central prediction from FONLL (green line) systemically underestimates the STAR data on $D^0$ production in p+p collisions at low-$p_T$, while the upper band of the prediction with $\mu_R=m_T/2$ is found to be closer to the measurements.
Moving to results for nuclear collisions, we can see that the transport approach, which include effects of non-equilibrium evolution, describes the experimental data on $p_T$ spectra at large $p_T>2$ GeV/$c$.
At low $p_T$, the calculations are still consistent with experiments within error bands but the central points of the band under-predict the experimental data.
We comment that the production of low-$p_T$ $D$ mesons in nuclear collisions is sensitive to the small-$x$ region of the nuclear parton distribution function (nPDF), which is less well-constrained than the parton distribution function of the proton. 
Despite the EPPS16 nPDF published its uncertainties; we have only shown calculations using the central fit of the EPPS16 nPDF.

Comparison of the high-$p_T$ part of the spectra between the hydrodynamic results and experimental data show that charm quarks have not reached complete equilibrium for $p_T>1$ GeV/$c$, since the spectra from experiments are still below the hydrodynamic results which assumes a fully equilibrated system of charm quarks that flow with the QGP fluid. The enhanced ratios of $\Lambda_c / D^0$ and $D_s/D^0$ observed at RHIC \cite{Zhou:2017ikn,Adam:2019hpq} for $Au+Au$ collisions and LHC \cite{Acharya:2018hre,Acharya:2018ckj} for $Pb+Pb$ collisions as compared with $p+p$ collisions may partially contribute to this $D^0$ suppression \cite{Greco:2003mm,Cao:2019iqs,He:2019vgs,Wang:2019fcg}.

Both the relativistic hydro and transport model can describe the experimental data on the elliptic flow of $D^0$ mesons within the error bars as shown in the right panel of Fig.~\ref{fig:v2 and pt spectrum}.  
One can see the mass ordering of the elliptic flow by comparing it to the elliptic flow of charged pions (blue) from the same hydrodynamic calculations.  
The hydrodynamic results on $v_2(p_T)$ at low $p_T$ are insensitive to the value of the freeze-out temperature.  

We stress that even though the hydrodynamics fails to describe the $p_T$ spectra of $D^0$ beyond $p_T>1$ GeV/$c$, its results on the elliptic flow agree with the experimental data and the transport model calculation (solid blue) for $p_T$ up to 4 GeV/$c$. 
This implies that the experimental data on the elliptic flow cannot provide a stringent constraint on the kinetic  equilibration of heavy quarks. The strong interaction between heavy quarks and the medium, however, can "drag" the heavy quarks along collective flow developed for the bulk medium even though the interaction might not drive the heavy quarks to full kinetic equilibration.

\section{Summary}

We have calculated the $D^0$ meson spectra and elliptic flow in Au+Au collisions at the RHIC energy within a relativistic viscous hydrodynamic model assuming that charm quarks initially produced through hard processes become fully kinetically equilibrated in the QGP. 
We neglect the thermal charm quark pair production in the QGP and use an effective charm chemical potential at the freeze-out which is tuned to describe the number of charm quarks produced in the initial hard processes. 
With the charm freeze-out temperature $T_{\rm frz}=137-150$ MeV, the hydrodynamic model can describe the experimental data on $D^0$ elliptic flow for $p_T<4$ GeV/$c$ as well as the elliptic flow of light quark hadrons. 
The mass ordering of the elliptic flow for pions and $D$ mesons in the experimental data is also observed for a freeze-out temperature between 137 and 150 MeV.
We have also compared the hydrodynamic results with that from a transport model with both energy loss and momentum diffusion, which can describe the observed elliptic flow of $D^0$ mesons at both low and high $p_T$.

The hydrodynamics, however, fails to describe the $p_T$ spectra of $D^0$ for $p_T>1$ GeV/$c$, significantly over-predicting the $D^0$ $p_T$ spectra at large $p_T$. 
The transport model, on the other hand, can describe well the $p_T$ spectra at large $p_T>$ 2 GeV/$c$ due to parton energy loss. 
But its central values under-predict the spectra at low $p_T$ due to the baseline spectra of the initial charm production through hard processes in p+p collisions. 

Our comparative study indicates that only the low-$p_T$ D mesons in the experiments might have reached kinetic equilibrium while charm quarks in the intermediate region of $p_T$ are partially thermalized. 
Parton energy loss and momentum diffusion in the transport model can describe well the non-equilibrium behavior of $D$ mesons spectra and elliptic flow at large $p_T$. 
The interaction of charm quarks in partial equilibrium with the medium in the intermediate $p_T$, however, can develop elliptic flow as large as the fully equilibrated charm quarks.

\begin{acknowledgments}
This work is supported by the National Key Research and Development Program of China under Grant No. 2020YFE0202002.
This work is supported in part by the National Science Foundation of China under Grant Nos. 11935007, 11221504, 11861131009, 12075098 and 11890714, and by the Director, Office of Energy Research, Office of High Energy and Nuclear Physics, Division of Nuclear Physics, of the U.S. Department of Energy (DOE) under grant No. DE- AC02-05CH11231, by the U.S. National Science Foundation under grant No. ACI-1550228 within JETSCAPE Collaboration, No. OAC- 2004571 within the X-SCAPE Collaboration by the UCB-CCNU Collaboration Grant. Computations are performed at Nuclear Science Computer Center at CCNU (NSC3) and the National Energy Research Scientific Computing Center (NERSC), a U.S. Department of Energy Office of Science User Facility operated under Contract No. DE-AC02- 05CH11231.

\end{acknowledgments}

%\printbibliography
%\bibliographystyle{unsrt}
%\bibliographystyle{aipnum4-1}
\bibliographystyle{apsrev4-1}
\bibliography{reference}

\begin{thebibliography}{61}%
\makeatletter
\providecommand \@ifxundefined [1]{%
 \@ifx{#1\undefined}
}%
\providecommand \@ifnum [1]{%
 \ifnum #1\expandafter \@firstoftwo
 \else \expandafter \@secondoftwo
 \fi
}%
\providecommand \@ifx [1]{%
 \ifx #1\expandafter \@firstoftwo
 \else \expandafter \@secondoftwo
 \fi
}%
\providecommand \natexlab [1]{#1}%
\providecommand \enquote  [1]{``#1''}%
\providecommand \bibnamefont  [1]{#1}%
\providecommand \bibfnamefont [1]{#1}%
\providecommand \citenamefont [1]{#1}%
\providecommand \href@noop [0]{\@secondoftwo}%
\providecommand \href [0]{\begingroup \@sanitize@url \@href}%
\providecommand \@href[1]{\@@startlink{#1}\@@href}%
\providecommand \@@href[1]{\endgroup#1\@@endlink}%
\providecommand \@sanitize@url [0]{\catcode `\\12\catcode `\$12\catcode
  `\&12\catcode `\#12\catcode `\^12\catcode `\_12\catcode `\%12\relax}%
\providecommand \@@startlink[1]{}%
\providecommand \@@endlink[0]{}%
\providecommand \url  [0]{\begingroup\@sanitize@url \@url }%
\providecommand \@url [1]{\endgroup\@href {#1}{\urlprefix }}%
\providecommand \urlprefix  [0]{URL }%
\providecommand \Eprint [0]{\href }%
\providecommand \doibase [0]{http://dx.doi.org/}%
\providecommand \selectlanguage [0]{\@gobble}%
\providecommand \bibinfo  [0]{\@secondoftwo}%
\providecommand \bibfield  [0]{\@secondoftwo}%
\providecommand \translation [1]{[#1]}%
\providecommand \BibitemOpen [0]{}%
\providecommand \bibitemStop [0]{}%
\providecommand \bibitemNoStop [0]{.\EOS\space}%
\providecommand \EOS [0]{\spacefactor3000\relax}%
\providecommand \BibitemShut  [1]{\csname bibitem#1\endcsname}%
\let\auto@bib@innerbib\@empty
%</preamble>
\bibitem [{\citenamefont {Gelman}\ \emph {et~al.}(2006)\citenamefont {Gelman},
  \citenamefont {Shuryak},\ and\ \citenamefont {Zahed}}]{Gelman:2006xw}%
  \BibitemOpen
  \bibfield  {author} {\bibinfo {author} {\bibfnamefont {B.~A.}\ \bibnamefont
  {Gelman}}, \bibinfo {author} {\bibfnamefont {E.~V.}\ \bibnamefont {Shuryak}},
  \ and\ \bibinfo {author} {\bibfnamefont {I.}~\bibnamefont {Zahed}},\ }\href
  {\doibase 10.1103/PhysRevC.74.044908} {\bibfield  {journal} {\bibinfo
  {journal} {Phys. Rev. C}\ }\textbf {\bibinfo {volume} {74}},\ \bibinfo
  {pages} {044908} (\bibinfo {year} {2006})},\ \Eprint
  {http://arxiv.org/abs/nucl-th/0601029} {arXiv:nucl-th/0601029} \BibitemShut
  {NoStop}%
\bibitem [{\citenamefont {Heinz}(2009)}]{Heinz:2008tv}%
  \BibitemOpen
  \bibfield  {author} {\bibinfo {author} {\bibfnamefont {U.~W.}\ \bibnamefont
  {Heinz}},\ }\href {\doibase 10.1088/1751-8113/42/21/214003} {\bibfield
  {journal} {\bibinfo  {journal} {J. Phys. A}\ }\textbf {\bibinfo {volume}
  {42}},\ \bibinfo {pages} {214003} (\bibinfo {year} {2009})},\ \Eprint
  {http://arxiv.org/abs/0810.5529} {arXiv:0810.5529 [nucl-th]} \BibitemShut
  {NoStop}%
\bibitem [{\citenamefont {Adler}\ \emph
  {et~al.}(2003{\natexlab{a}})\citenamefont {Adler} \emph
  {et~al.}}]{Adler:2003kt}%
  \BibitemOpen
  \bibfield  {author} {\bibinfo {author} {\bibfnamefont {S.}~\bibnamefont
  {Adler}} \emph {et~al.} (\bibinfo {collaboration} {PHENIX}),\ }\href
  {\doibase 10.1103/PhysRevLett.91.182301} {\bibfield  {journal} {\bibinfo
  {journal} {Phys. Rev. Lett.}\ }\textbf {\bibinfo {volume} {91}},\ \bibinfo
  {pages} {182301} (\bibinfo {year} {2003}{\natexlab{a}})},\ \Eprint
  {http://arxiv.org/abs/nucl-ex/0305013} {arXiv:nucl-ex/0305013} \BibitemShut
  {NoStop}%
\bibitem [{\citenamefont {Adams}\ \emph {et~al.}(2004)\citenamefont {Adams}
  \emph {et~al.}}]{Adams:2003am}%
  \BibitemOpen
  \bibfield  {author} {\bibinfo {author} {\bibfnamefont {J.}~\bibnamefont
  {Adams}} \emph {et~al.} (\bibinfo {collaboration} {STAR}),\ }\href {\doibase
  10.1103/PhysRevLett.92.052302} {\bibfield  {journal} {\bibinfo  {journal}
  {Phys. Rev. Lett.}\ }\textbf {\bibinfo {volume} {92}},\ \bibinfo {pages}
  {052302} (\bibinfo {year} {2004})},\ \Eprint
  {http://arxiv.org/abs/nucl-ex/0306007} {arXiv:nucl-ex/0306007} \BibitemShut
  {NoStop}%
\bibitem [{\citenamefont {Abelev}\ \emph {et~al.}(2015)\citenamefont {Abelev}
  \emph {et~al.}}]{Abelev:2014pua}%
  \BibitemOpen
  \bibfield  {author} {\bibinfo {author} {\bibfnamefont {B.~B.}\ \bibnamefont
  {Abelev}} \emph {et~al.} (\bibinfo {collaboration} {ALICE}),\ }\href
  {\doibase 10.1007/JHEP06(2015)190} {\bibfield  {journal} {\bibinfo  {journal}
  {JHEP}\ }\textbf {\bibinfo {volume} {06}},\ \bibinfo {pages} {190} (\bibinfo
  {year} {2015})},\ \Eprint {http://arxiv.org/abs/1405.4632} {arXiv:1405.4632
  [nucl-ex]} \BibitemShut {NoStop}%
\bibitem [{\citenamefont {Adler}\ \emph
  {et~al.}(2003{\natexlab{b}})\citenamefont {Adler} \emph
  {et~al.}}]{Adler:2002tq}%
  \BibitemOpen
  \bibfield  {author} {\bibinfo {author} {\bibfnamefont {C.}~\bibnamefont
  {Adler}} \emph {et~al.} (\bibinfo {collaboration} {STAR}),\ }\href {\doibase
  10.1103/PhysRevLett.90.082302} {\bibfield  {journal} {\bibinfo  {journal}
  {Phys. Rev. Lett.}\ }\textbf {\bibinfo {volume} {90}},\ \bibinfo {pages}
  {082302} (\bibinfo {year} {2003}{\natexlab{b}})},\ \Eprint
  {http://arxiv.org/abs/nucl-ex/0210033} {arXiv:nucl-ex/0210033} \BibitemShut
  {NoStop}%
\bibitem [{\citenamefont {Adcox}\ \emph {et~al.}(2002)\citenamefont {Adcox}
  \emph {et~al.}}]{Adcox:2001jp}%
  \BibitemOpen
  \bibfield  {author} {\bibinfo {author} {\bibfnamefont {K.}~\bibnamefont
  {Adcox}} \emph {et~al.} (\bibinfo {collaboration} {PHENIX}),\ }\href
  {\doibase 10.1103/PhysRevLett.88.022301} {\bibfield  {journal} {\bibinfo
  {journal} {Phys. Rev. Lett.}\ }\textbf {\bibinfo {volume} {88}},\ \bibinfo
  {pages} {022301} (\bibinfo {year} {2002})},\ \Eprint
  {http://arxiv.org/abs/nucl-ex/0109003} {arXiv:nucl-ex/0109003} \BibitemShut
  {NoStop}%
\bibitem [{\citenamefont {Adler}\ \emph
  {et~al.}(2003{\natexlab{c}})\citenamefont {Adler} \emph
  {et~al.}}]{Adler:2003qi}%
  \BibitemOpen
  \bibfield  {author} {\bibinfo {author} {\bibfnamefont {S.}~\bibnamefont
  {Adler}} \emph {et~al.} (\bibinfo {collaboration} {PHENIX}),\ }\href
  {\doibase 10.1103/PhysRevLett.91.072301} {\bibfield  {journal} {\bibinfo
  {journal} {Phys. Rev. Lett.}\ }\textbf {\bibinfo {volume} {91}},\ \bibinfo
  {pages} {072301} (\bibinfo {year} {2003}{\natexlab{c}})},\ \Eprint
  {http://arxiv.org/abs/nucl-ex/0304022} {arXiv:nucl-ex/0304022} \BibitemShut
  {NoStop}%
\bibitem [{\citenamefont {Adler}\ \emph {et~al.}(2002)\citenamefont {Adler}
  \emph {et~al.}}]{Adler:2002xw}%
  \BibitemOpen
  \bibfield  {author} {\bibinfo {author} {\bibfnamefont {C.}~\bibnamefont
  {Adler}} \emph {et~al.} (\bibinfo {collaboration} {STAR}),\ }\href {\doibase
  10.1103/PhysRevLett.89.202301} {\bibfield  {journal} {\bibinfo  {journal}
  {Phys. Rev. Lett.}\ }\textbf {\bibinfo {volume} {89}},\ \bibinfo {pages}
  {202301} (\bibinfo {year} {2002})},\ \Eprint
  {http://arxiv.org/abs/nucl-ex/0206011} {arXiv:nucl-ex/0206011} \BibitemShut
  {NoStop}%
\bibitem [{\citenamefont {Chatrchyan}\ \emph {et~al.}(2012)\citenamefont
  {Chatrchyan} \emph {et~al.}}]{CMS:2012aa}%
  \BibitemOpen
  \bibfield  {author} {\bibinfo {author} {\bibfnamefont {S.}~\bibnamefont
  {Chatrchyan}} \emph {et~al.} (\bibinfo {collaboration} {CMS}),\ }\href
  {\doibase 10.1140/epjc/s10052-012-1945-x} {\bibfield  {journal} {\bibinfo
  {journal} {Eur. Phys. J. C}\ }\textbf {\bibinfo {volume} {72}},\ \bibinfo
  {pages} {1945} (\bibinfo {year} {2012})},\ \Eprint
  {http://arxiv.org/abs/1202.2554} {arXiv:1202.2554 [nucl-ex]} \BibitemShut
  {NoStop}%
\bibitem [{\citenamefont {Adamczyk}\ \emph {et~al.}(2014)\citenamefont
  {Adamczyk} \emph {et~al.}}]{Adamczyk:2014uip}%
  \BibitemOpen
  \bibfield  {author} {\bibinfo {author} {\bibfnamefont {L.}~\bibnamefont
  {Adamczyk}} \emph {et~al.} (\bibinfo {collaboration} {STAR}),\ }\href
  {\doibase 10.1103/PhysRevLett.113.142301} {\bibfield  {journal} {\bibinfo
  {journal} {Phys. Rev. Lett.}\ }\textbf {\bibinfo {volume} {113}},\ \bibinfo
  {pages} {142301} (\bibinfo {year} {2014})},\ \bibinfo {note} {[Erratum:
  Phys.Rev.Lett. 121, 229901 (2018)]},\ \Eprint
  {http://arxiv.org/abs/1404.6185} {arXiv:1404.6185 [nucl-ex]} \BibitemShut
  {NoStop}%
\bibitem [{\citenamefont {Adam}\ \emph {et~al.}(2019)\citenamefont {Adam} \emph
  {et~al.}}]{Adam:2018inb}%
  \BibitemOpen
  \bibfield  {author} {\bibinfo {author} {\bibfnamefont {J.}~\bibnamefont
  {Adam}} \emph {et~al.} (\bibinfo {collaboration} {STAR}),\ }\href {\doibase
  10.1103/PhysRevC.99.034908} {\bibfield  {journal} {\bibinfo  {journal} {Phys.
  Rev. C}\ }\textbf {\bibinfo {volume} {99}},\ \bibinfo {pages} {034908}
  (\bibinfo {year} {2019})},\ \Eprint {http://arxiv.org/abs/1812.10224}
  {arXiv:1812.10224 [nucl-ex]} \BibitemShut {NoStop}%
\bibitem [{\citenamefont {Adamczyk}\ \emph {et~al.}(2017)\citenamefont
  {Adamczyk} \emph {et~al.}}]{Adamczyk:2017xur}%
  \BibitemOpen
  \bibfield  {author} {\bibinfo {author} {\bibfnamefont {L.}~\bibnamefont
  {Adamczyk}} \emph {et~al.} (\bibinfo {collaboration} {STAR}),\ }\href
  {\doibase 10.1103/PhysRevLett.118.212301} {\bibfield  {journal} {\bibinfo
  {journal} {Phys. Rev. Lett.}\ }\textbf {\bibinfo {volume} {118}},\ \bibinfo
  {pages} {212301} (\bibinfo {year} {2017})},\ \Eprint
  {http://arxiv.org/abs/1701.06060} {arXiv:1701.06060 [nucl-ex]} \BibitemShut
  {NoStop}%
%%CITATION = ARXIV:1701.06060;%%
\bibitem [{\citenamefont {Abelev}\ \emph {et~al.}(2012)\citenamefont {Abelev}
  \emph {et~al.}}]{ALICE:2012ab}%
  \BibitemOpen
  \bibfield  {author} {\bibinfo {author} {\bibfnamefont {B.}~\bibnamefont
  {Abelev}} \emph {et~al.} (\bibinfo {collaboration} {ALICE}),\ }\href
  {\doibase 10.1007/JHEP09(2012)112} {\bibfield  {journal} {\bibinfo  {journal}
  {JHEP}\ }\textbf {\bibinfo {volume} {09}},\ \bibinfo {pages} {112} (\bibinfo
  {year} {2012})},\ \Eprint {http://arxiv.org/abs/1203.2160} {arXiv:1203.2160
  [nucl-ex]} \BibitemShut {NoStop}%
\bibitem [{\citenamefont {Abelev}\ \emph {et~al.}(2013)\citenamefont {Abelev}
  \emph {et~al.}}]{Abelev:2013lca}%
  \BibitemOpen
  \bibfield  {author} {\bibinfo {author} {\bibfnamefont {B.}~\bibnamefont
  {Abelev}} \emph {et~al.} (\bibinfo {collaboration} {ALICE}),\ }\href
  {\doibase 10.1103/PhysRevLett.111.102301} {\bibfield  {journal} {\bibinfo
  {journal} {Phys. Rev. Lett.}\ }\textbf {\bibinfo {volume} {111}},\ \bibinfo
  {pages} {102301} (\bibinfo {year} {2013})},\ \Eprint
  {http://arxiv.org/abs/1305.2707} {arXiv:1305.2707 [nucl-ex]} \BibitemShut
  {NoStop}%
\bibitem [{\citenamefont {Sirunyan}\ \emph
  {et~al.}(2018{\natexlab{a}})\citenamefont {Sirunyan} \emph
  {et~al.}}]{Sirunyan:2017xss}%
  \BibitemOpen
  \bibfield  {author} {\bibinfo {author} {\bibfnamefont {A.~M.}\ \bibnamefont
  {Sirunyan}} \emph {et~al.} (\bibinfo {collaboration} {CMS}),\ }\href
  {\doibase 10.1016/j.physletb.2018.05.074} {\bibfield  {journal} {\bibinfo
  {journal} {Phys. Lett. B}\ }\textbf {\bibinfo {volume} {782}},\ \bibinfo
  {pages} {474} (\bibinfo {year} {2018}{\natexlab{a}})},\ \Eprint
  {http://arxiv.org/abs/1708.04962} {arXiv:1708.04962 [nucl-ex]} \BibitemShut
  {NoStop}%
\bibitem [{\citenamefont {Sirunyan}\ \emph
  {et~al.}(2018{\natexlab{b}})\citenamefont {Sirunyan} \emph
  {et~al.}}]{Sirunyan:2017plt}%
  \BibitemOpen
  \bibfield  {author} {\bibinfo {author} {\bibfnamefont {A.~M.}\ \bibnamefont
  {Sirunyan}} \emph {et~al.} (\bibinfo {collaboration} {CMS}),\ }\href
  {\doibase 10.1103/PhysRevLett.120.202301} {\bibfield  {journal} {\bibinfo
  {journal} {Phys. Rev. Lett.}\ }\textbf {\bibinfo {volume} {120}},\ \bibinfo
  {pages} {202301} (\bibinfo {year} {2018}{\natexlab{b}})},\ \Eprint
  {http://arxiv.org/abs/1708.03497} {arXiv:1708.03497 [nucl-ex]} \BibitemShut
  {NoStop}%
\bibitem [{\citenamefont {Sirunyan}\ \emph
  {et~al.}(2018{\natexlab{c}})\citenamefont {Sirunyan} \emph
  {et~al.}}]{Sirunyan:2018toe}%
  \BibitemOpen
  \bibfield  {author} {\bibinfo {author} {\bibfnamefont {A.~M.}\ \bibnamefont
  {Sirunyan}} \emph {et~al.} (\bibinfo {collaboration} {CMS}),\ }\href
  {\doibase 10.1103/PhysRevLett.121.082301} {\bibfield  {journal} {\bibinfo
  {journal} {Phys. Rev. Lett.}\ }\textbf {\bibinfo {volume} {121}},\ \bibinfo
  {pages} {082301} (\bibinfo {year} {2018}{\natexlab{c}})},\ \Eprint
  {http://arxiv.org/abs/1804.09767} {arXiv:1804.09767 [hep-ex]} \BibitemShut
  {NoStop}%
\bibitem [{\citenamefont {Wang}\ and\ \citenamefont
  {Uhlenbeck}(1945)}]{RevModPhys.17.323}%
  \BibitemOpen
  \bibfield  {author} {\bibinfo {author} {\bibfnamefont {M.~C.}\ \bibnamefont
  {Wang}}\ and\ \bibinfo {author} {\bibfnamefont {G.~E.}\ \bibnamefont
  {Uhlenbeck}},\ }\href {\doibase 10.1103/RevModPhys.17.323} {\bibfield
  {journal} {\bibinfo  {journal} {Rev. Mod. Phys.}\ }\textbf {\bibinfo {volume}
  {17}},\ \bibinfo {pages} {323} (\bibinfo {year} {1945})}\BibitemShut
  {NoStop}%
\bibitem [{\citenamefont {Rapp}\ and\ \citenamefont {van
  Hees}(2010)}]{Rapp:2009my}%
  \BibitemOpen
  \bibfield  {author} {\bibinfo {author} {\bibfnamefont {R.}~\bibnamefont
  {Rapp}}\ and\ \bibinfo {author} {\bibfnamefont {H.}~\bibnamefont {van Hees}}\
  }(\bibinfo {year} {2010})\ pp.\ \bibinfo {pages} {111--206},\ \Eprint
  {http://arxiv.org/abs/0903.1096} {arXiv:0903.1096 [hep-ph]} \BibitemShut
  {NoStop}%
\bibitem [{\citenamefont {Moore}\ and\ \citenamefont
  {Teaney}(2005)}]{Moore:2004tg}%
  \BibitemOpen
  \bibfield  {author} {\bibinfo {author} {\bibfnamefont {G.~D.}\ \bibnamefont
  {Moore}}\ and\ \bibinfo {author} {\bibfnamefont {D.}~\bibnamefont {Teaney}},\
  }\href {\doibase 10.1103/PhysRevC.71.064904} {\bibfield  {journal} {\bibinfo
  {journal} {Phys. Rev. C}\ }\textbf {\bibinfo {volume} {71}},\ \bibinfo
  {pages} {064904} (\bibinfo {year} {2005})},\ \Eprint
  {http://arxiv.org/abs/hep-ph/0412346} {arXiv:hep-ph/0412346} \BibitemShut
  {NoStop}%
\bibitem [{\citenamefont {Cao}\ \emph {et~al.}(2016)\citenamefont {Cao},
  \citenamefont {Luo}, \citenamefont {Qin},\ and\ \citenamefont
  {Wang}}]{Cao:2016gvr}%
  \BibitemOpen
  \bibfield  {author} {\bibinfo {author} {\bibfnamefont {S.}~\bibnamefont
  {Cao}}, \bibinfo {author} {\bibfnamefont {T.}~\bibnamefont {Luo}}, \bibinfo
  {author} {\bibfnamefont {G.-Y.}\ \bibnamefont {Qin}}, \ and\ \bibinfo
  {author} {\bibfnamefont {X.-N.}\ \bibnamefont {Wang}},\ }\href {\doibase
  10.1103/PhysRevC.94.014909} {\bibfield  {journal} {\bibinfo  {journal} {Phys.
  Rev. C}\ }\textbf {\bibinfo {volume} {94}},\ \bibinfo {pages} {014909}
  (\bibinfo {year} {2016})},\ \Eprint {http://arxiv.org/abs/1605.06447}
  {arXiv:1605.06447 [nucl-th]} \BibitemShut {NoStop}%
\bibitem [{\citenamefont {Nahrgang}\ \emph {et~al.}(2016)\citenamefont
  {Nahrgang}, \citenamefont {Aichelin}, \citenamefont {Gossiaux},\ and\
  \citenamefont {Werner}}]{Nahrgang:2016lst}%
  \BibitemOpen
  \bibfield  {author} {\bibinfo {author} {\bibfnamefont {M.}~\bibnamefont
  {Nahrgang}}, \bibinfo {author} {\bibfnamefont {J.}~\bibnamefont {Aichelin}},
  \bibinfo {author} {\bibfnamefont {P.~B.}\ \bibnamefont {Gossiaux}}, \ and\
  \bibinfo {author} {\bibfnamefont {K.}~\bibnamefont {Werner}},\ }\href
  {\doibase 10.1103/PhysRevC.93.044909} {\bibfield  {journal} {\bibinfo
  {journal} {Phys. Rev. C}\ }\textbf {\bibinfo {volume} {93}},\ \bibinfo
  {pages} {044909} (\bibinfo {year} {2016})},\ \Eprint
  {http://arxiv.org/abs/1602.03544} {arXiv:1602.03544 [nucl-th]} \BibitemShut
  {NoStop}%
\bibitem [{\citenamefont {Ke}\ \emph {et~al.}(2018)\citenamefont {Ke},
  \citenamefont {Xu},\ and\ \citenamefont {Bass}}]{WeiYao2018}%
  \BibitemOpen
  \bibfield  {author} {\bibinfo {author} {\bibfnamefont {W.}~\bibnamefont
  {Ke}}, \bibinfo {author} {\bibfnamefont {Y.}~\bibnamefont {Xu}}, \ and\
  \bibinfo {author} {\bibfnamefont {S.~A.}\ \bibnamefont {Bass}},\ }\href
  {\doibase 10.1103/PhysRevC.98.064901} {\bibfield  {journal} {\bibinfo
  {journal} {Phys. Rev. C}\ }\textbf {\bibinfo {volume} {98}},\ \bibinfo
  {pages} {064901} (\bibinfo {year} {2018})}\BibitemShut {NoStop}%
\bibitem [{\citenamefont {Cao}\ \emph {et~al.}(2019)\citenamefont {Cao} \emph
  {et~al.}}]{Cao:2018ews}%
  \BibitemOpen
  \bibfield  {author} {\bibinfo {author} {\bibfnamefont {S.}~\bibnamefont
  {Cao}} \emph {et~al.},\ }\href {\doibase 10.1103/PhysRevC.99.054907}
  {\bibfield  {journal} {\bibinfo  {journal} {Phys. Rev. C}\ }\textbf {\bibinfo
  {volume} {99}},\ \bibinfo {pages} {054907} (\bibinfo {year} {2019})},\
  \Eprint {http://arxiv.org/abs/1809.07894} {arXiv:1809.07894 [nucl-th]}
  \BibitemShut {NoStop}%
\bibitem [{\citenamefont {Beraudo}\ \emph {et~al.}(2018)\citenamefont {Beraudo}
  \emph {et~al.}}]{Rapp:2018qla}%
  \BibitemOpen
  \bibfield  {author} {\bibinfo {author} {\bibfnamefont {A.}~\bibnamefont
  {Beraudo}} \emph {et~al.},\ }\href {\doibase 10.1016/j.nuclphysa.2018.09.002}
  {\bibfield  {journal} {\bibinfo  {journal} {Nucl. Phys. A}\ }\textbf
  {\bibinfo {volume} {979}},\ \bibinfo {pages} {21} (\bibinfo {year} {2018})},\
  \Eprint {http://arxiv.org/abs/1803.03824} {arXiv:1803.03824 [nucl-th]}
  \BibitemShut {NoStop}%
\bibitem [{\citenamefont {Banerjee}\ \emph {et~al.}(2012)\citenamefont
  {Banerjee}, \citenamefont {Datta}, \citenamefont {Gavai},\ and\ \citenamefont
  {Majumdar}}]{Banerjee:2011ra}%
  \BibitemOpen
  \bibfield  {author} {\bibinfo {author} {\bibfnamefont {D.}~\bibnamefont
  {Banerjee}}, \bibinfo {author} {\bibfnamefont {S.}~\bibnamefont {Datta}},
  \bibinfo {author} {\bibfnamefont {R.}~\bibnamefont {Gavai}}, \ and\ \bibinfo
  {author} {\bibfnamefont {P.}~\bibnamefont {Majumdar}},\ }\href {\doibase
  10.1103/PhysRevD.85.014510} {\bibfield  {journal} {\bibinfo  {journal} {Phys.
  Rev. D}\ }\textbf {\bibinfo {volume} {85}},\ \bibinfo {pages} {014510}
  (\bibinfo {year} {2012})},\ \Eprint {http://arxiv.org/abs/1109.5738}
  {arXiv:1109.5738 [hep-lat]} \BibitemShut {NoStop}%
\bibitem [{\citenamefont {Ding}\ \emph {et~al.}(2012)\citenamefont {Ding},
  \citenamefont {Francis}, \citenamefont {Kaczmarek}, \citenamefont {Karsch},
  \citenamefont {Satz},\ and\ \citenamefont {Soeldner}}]{Ding:2012sp}%
  \BibitemOpen
  \bibfield  {author} {\bibinfo {author} {\bibfnamefont {H.}~\bibnamefont
  {Ding}}, \bibinfo {author} {\bibfnamefont {A.}~\bibnamefont {Francis}},
  \bibinfo {author} {\bibfnamefont {O.}~\bibnamefont {Kaczmarek}}, \bibinfo
  {author} {\bibfnamefont {F.}~\bibnamefont {Karsch}}, \bibinfo {author}
  {\bibfnamefont {H.}~\bibnamefont {Satz}}, \ and\ \bibinfo {author}
  {\bibfnamefont {W.}~\bibnamefont {Soeldner}},\ }\href {\doibase
  10.1103/PhysRevD.86.014509} {\bibfield  {journal} {\bibinfo  {journal} {Phys.
  Rev. D}\ }\textbf {\bibinfo {volume} {86}},\ \bibinfo {pages} {014509}
  (\bibinfo {year} {2012})},\ \Eprint {http://arxiv.org/abs/1204.4945}
  {arXiv:1204.4945 [hep-lat]} \BibitemShut {NoStop}%
\bibitem [{\citenamefont {Francis}\ \emph {et~al.}(2015)\citenamefont
  {Francis}, \citenamefont {Kaczmarek}, \citenamefont {Laine}, \citenamefont
  {Neuhaus},\ and\ \citenamefont {Ohno}}]{Francis:2015daa}%
  \BibitemOpen
  \bibfield  {author} {\bibinfo {author} {\bibfnamefont {A.}~\bibnamefont
  {Francis}}, \bibinfo {author} {\bibfnamefont {O.}~\bibnamefont {Kaczmarek}},
  \bibinfo {author} {\bibfnamefont {M.}~\bibnamefont {Laine}}, \bibinfo
  {author} {\bibfnamefont {T.}~\bibnamefont {Neuhaus}}, \ and\ \bibinfo
  {author} {\bibfnamefont {H.}~\bibnamefont {Ohno}},\ }\href {\doibase
  10.1103/PhysRevD.92.116003} {\bibfield  {journal} {\bibinfo  {journal} {Phys.
  Rev. D}\ }\textbf {\bibinfo {volume} {92}},\ \bibinfo {pages} {116003}
  (\bibinfo {year} {2015})},\ \Eprint {http://arxiv.org/abs/1508.04543}
  {arXiv:1508.04543 [hep-lat]} \BibitemShut {NoStop}%
\bibitem [{\citenamefont {Xu}\ \emph {et~al.}(2018)\citenamefont {Xu},
  \citenamefont {Bernhard}, \citenamefont {Bass}, \citenamefont {Nahrgang},\
  and\ \citenamefont {Cao}}]{Xu:2017obm}%
  \BibitemOpen
  \bibfield  {author} {\bibinfo {author} {\bibfnamefont {Y.}~\bibnamefont
  {Xu}}, \bibinfo {author} {\bibfnamefont {J.~E.}\ \bibnamefont {Bernhard}},
  \bibinfo {author} {\bibfnamefont {S.~A.}\ \bibnamefont {Bass}}, \bibinfo
  {author} {\bibfnamefont {M.}~\bibnamefont {Nahrgang}}, \ and\ \bibinfo
  {author} {\bibfnamefont {S.}~\bibnamefont {Cao}},\ }\href {\doibase
  10.1103/PhysRevC.97.014907} {\bibfield  {journal} {\bibinfo  {journal} {Phys.
  Rev. C}\ }\textbf {\bibinfo {volume} {97}},\ \bibinfo {pages} {014907}
  (\bibinfo {year} {2018})},\ \Eprint {http://arxiv.org/abs/1710.00807}
  {arXiv:1710.00807 [nucl-th]} \BibitemShut {NoStop}%
\bibitem [{\citenamefont {Cooper}\ and\ \citenamefont
  {Frye}(1974)}]{Cooper:1974mv}%
  \BibitemOpen
  \bibfield  {author} {\bibinfo {author} {\bibfnamefont {F.}~\bibnamefont
  {Cooper}}\ and\ \bibinfo {author} {\bibfnamefont {G.}~\bibnamefont {Frye}},\
  }\href {\doibase 10.1103/PhysRevD.10.186} {\bibfield  {journal} {\bibinfo
  {journal} {Phys. Rev. D}\ }\textbf {\bibinfo {volume} {10}},\ \bibinfo
  {pages} {186} (\bibinfo {year} {1974})}\BibitemShut {NoStop}%
\bibitem [{\citenamefont {Pang}\ \emph {et~al.}(2018)\citenamefont {Pang},
  \citenamefont {Petersen},\ and\ \citenamefont {Wang}}]{Pang:2018zzo}%
  \BibitemOpen
  \bibfield  {author} {\bibinfo {author} {\bibfnamefont {L.-G.}\ \bibnamefont
  {Pang}}, \bibinfo {author} {\bibfnamefont {H.}~\bibnamefont {Petersen}}, \
  and\ \bibinfo {author} {\bibfnamefont {X.-N.}\ \bibnamefont {Wang}},\ }\href
  {\doibase 10.1103/PhysRevC.97.064918} {\bibfield  {journal} {\bibinfo
  {journal} {Phys. Rev. C}\ }\textbf {\bibinfo {volume} {97}},\ \bibinfo
  {pages} {064918} (\bibinfo {year} {2018})},\ \Eprint
  {http://arxiv.org/abs/1802.04449} {arXiv:1802.04449 [nucl-th]} \BibitemShut
  {NoStop}%
\bibitem [{\citenamefont {Collins}\ \emph {et~al.}(1986)\citenamefont
  {Collins}, \citenamefont {Soper},\ and\ \citenamefont
  {Sterman}}]{Collins:1985gm}%
  \BibitemOpen
  \bibfield  {author} {\bibinfo {author} {\bibfnamefont {J.~C.}\ \bibnamefont
  {Collins}}, \bibinfo {author} {\bibfnamefont {D.~E.}\ \bibnamefont {Soper}},
  \ and\ \bibinfo {author} {\bibfnamefont {G.~F.}\ \bibnamefont {Sterman}},\
  }\href {\doibase 10.1016/0550-3213(86)90026-X} {\bibfield  {journal}
  {\bibinfo  {journal} {Nucl. Phys. B}\ }\textbf {\bibinfo {volume} {263}},\
  \bibinfo {pages} {37} (\bibinfo {year} {1986})}\BibitemShut {NoStop}%
\bibitem [{\citenamefont {Braaten}\ \emph {et~al.}(1995)\citenamefont
  {Braaten}, \citenamefont {Cheung}, \citenamefont {Fleming},\ and\
  \citenamefont {Yuan}}]{Braaten:1994bz}%
  \BibitemOpen
  \bibfield  {author} {\bibinfo {author} {\bibfnamefont {E.}~\bibnamefont
  {Braaten}}, \bibinfo {author} {\bibfnamefont {K.-m.}\ \bibnamefont {Cheung}},
  \bibinfo {author} {\bibfnamefont {S.}~\bibnamefont {Fleming}}, \ and\
  \bibinfo {author} {\bibfnamefont {T.~C.}\ \bibnamefont {Yuan}},\ }\href
  {\doibase 10.1103/PhysRevD.51.4819} {\bibfield  {journal} {\bibinfo
  {journal} {Phys. Rev. D}\ }\textbf {\bibinfo {volume} {51}},\ \bibinfo
  {pages} {4819} (\bibinfo {year} {1995})},\ \Eprint
  {http://arxiv.org/abs/hep-ph/9409316} {arXiv:hep-ph/9409316} \BibitemShut
  {NoStop}%
\bibitem [{\citenamefont {Cacciari}\ and\ \citenamefont
  {Nason}(2003)}]{Cacciari:2003zu}%
  \BibitemOpen
  \bibfield  {author} {\bibinfo {author} {\bibfnamefont {M.}~\bibnamefont
  {Cacciari}}\ and\ \bibinfo {author} {\bibfnamefont {P.}~\bibnamefont
  {Nason}},\ }\href {\doibase 10.1088/1126-6708/2003/09/006} {\bibfield
  {journal} {\bibinfo  {journal} {JHEP}\ }\textbf {\bibinfo {volume} {09}},\
  \bibinfo {pages} {006} (\bibinfo {year} {2003})},\ \Eprint
  {http://arxiv.org/abs/hep-ph/0306212} {arXiv:hep-ph/0306212} \BibitemShut
  {NoStop}%
\bibitem [{\citenamefont {Cacciari}\ \emph {et~al.}(2012)\citenamefont
  {Cacciari}, \citenamefont {Frixione}, \citenamefont {Houdeau}, \citenamefont
  {Mangano}, \citenamefont {Nason},\ and\ \citenamefont
  {Ridolfi}}]{Cacciari:2012ny}%
  \BibitemOpen
  \bibfield  {author} {\bibinfo {author} {\bibfnamefont {M.}~\bibnamefont
  {Cacciari}}, \bibinfo {author} {\bibfnamefont {S.}~\bibnamefont {Frixione}},
  \bibinfo {author} {\bibfnamefont {N.}~\bibnamefont {Houdeau}}, \bibinfo
  {author} {\bibfnamefont {M.~L.}\ \bibnamefont {Mangano}}, \bibinfo {author}
  {\bibfnamefont {P.}~\bibnamefont {Nason}}, \ and\ \bibinfo {author}
  {\bibfnamefont {G.}~\bibnamefont {Ridolfi}},\ }\href {\doibase
  10.1007/JHEP10(2012)137} {\bibfield  {journal} {\bibinfo  {journal} {JHEP}\
  }\textbf {\bibinfo {volume} {10}},\ \bibinfo {pages} {137} (\bibinfo {year}
  {2012})},\ \Eprint {http://arxiv.org/abs/1205.6344} {arXiv:1205.6344
  [hep-ph]} \BibitemShut {NoStop}%
\bibitem [{\citenamefont {Rapp}\ and\ \citenamefont
  {Shuryak}(2003)}]{Rapp:2003wn}%
  \BibitemOpen
  \bibfield  {author} {\bibinfo {author} {\bibfnamefont {R.}~\bibnamefont
  {Rapp}}\ and\ \bibinfo {author} {\bibfnamefont {E.}~\bibnamefont {Shuryak}},\
  }\href {\doibase 10.1103/PhysRevD.67.074036} {\bibfield  {journal} {\bibinfo
  {journal} {Phys. Rev. D}\ }\textbf {\bibinfo {volume} {67}},\ \bibinfo
  {pages} {074036} (\bibinfo {year} {2003})},\ \Eprint
  {http://arxiv.org/abs/hep-ph/0301245} {arXiv:hep-ph/0301245} \BibitemShut
  {NoStop}%
\bibitem [{\citenamefont {Patrignani}\ \emph {et~al.}(2016)\citenamefont
  {Patrignani} \emph {et~al.}}]{Patrignani:2016xqp}%
  \BibitemOpen
  \bibfield  {author} {\bibinfo {author} {\bibfnamefont {C.}~\bibnamefont
  {Patrignani}} \emph {et~al.} (\bibinfo {collaboration} {Particle Data
  Group}),\ }\href {\doibase 10.1088/1674-1137/40/10/100001} {\bibfield
  {journal} {\bibinfo  {journal} {Chin. Phys. C}\ }\textbf {\bibinfo {volume}
  {40}},\ \bibinfo {pages} {100001} (\bibinfo {year} {2016})}\BibitemShut
  {NoStop}%
\bibitem [{\citenamefont {Moreland}\ \emph {et~al.}(2015)\citenamefont
  {Moreland}, \citenamefont {Bernhard},\ and\ \citenamefont
  {Bass}}]{Moreland:2014oya}%
  \BibitemOpen
  \bibfield  {author} {\bibinfo {author} {\bibfnamefont {J.~S.}\ \bibnamefont
  {Moreland}}, \bibinfo {author} {\bibfnamefont {J.~E.}\ \bibnamefont
  {Bernhard}}, \ and\ \bibinfo {author} {\bibfnamefont {S.~A.}\ \bibnamefont
  {Bass}},\ }\href {\doibase 10.1103/PhysRevC.92.011901} {\bibfield  {journal}
  {\bibinfo  {journal} {Phys. Rev. C}\ }\textbf {\bibinfo {volume} {92}},\
  \bibinfo {pages} {011901} (\bibinfo {year} {2015})},\ \Eprint
  {http://arxiv.org/abs/1412.4708} {arXiv:1412.4708 [nucl-th]} \BibitemShut
  {NoStop}%
\bibitem [{\citenamefont {Bernhard}\ \emph {et~al.}(2016)\citenamefont
  {Bernhard}, \citenamefont {Moreland}, \citenamefont {Bass}, \citenamefont
  {Liu},\ and\ \citenamefont {Heinz}}]{Bernhard:2016tnd}%
  \BibitemOpen
  \bibfield  {author} {\bibinfo {author} {\bibfnamefont {J.~E.}\ \bibnamefont
  {Bernhard}}, \bibinfo {author} {\bibfnamefont {J.~S.}\ \bibnamefont
  {Moreland}}, \bibinfo {author} {\bibfnamefont {S.~A.}\ \bibnamefont {Bass}},
  \bibinfo {author} {\bibfnamefont {J.}~\bibnamefont {Liu}}, \ and\ \bibinfo
  {author} {\bibfnamefont {U.}~\bibnamefont {Heinz}},\ }\href {\doibase
  10.1103/PhysRevC.94.024907} {\bibfield  {journal} {\bibinfo  {journal} {Phys.
  Rev. C}\ }\textbf {\bibinfo {volume} {94}},\ \bibinfo {pages} {024907}
  (\bibinfo {year} {2016})},\ \Eprint {http://arxiv.org/abs/1605.03954}
  {arXiv:1605.03954 [nucl-th]} \BibitemShut {NoStop}%
\bibitem [{\citenamefont {Huovinen}\ and\ \citenamefont
  {Petreczky}(2010)}]{Huovinen:2009yb}%
  \BibitemOpen
  \bibfield  {author} {\bibinfo {author} {\bibfnamefont {P.}~\bibnamefont
  {Huovinen}}\ and\ \bibinfo {author} {\bibfnamefont {P.}~\bibnamefont
  {Petreczky}},\ }\href {\doibase 10.1016/j.nuclphysa.2010.02.015} {\bibfield
  {journal} {\bibinfo  {journal} {Nucl. Phys. A}\ }\textbf {\bibinfo {volume}
  {837}},\ \bibinfo {pages} {26} (\bibinfo {year} {2010})},\ \Eprint
  {http://arxiv.org/abs/0912.2541} {arXiv:0912.2541 [hep-ph]} \BibitemShut
  {NoStop}%
\bibitem [{\citenamefont {Cacciari}\ \emph {et~al.}(1998)\citenamefont
  {Cacciari}, \citenamefont {Greco},\ and\ \citenamefont
  {Nason}}]{Cacciari:1998it}%
  \BibitemOpen
  \bibfield  {author} {\bibinfo {author} {\bibfnamefont {M.}~\bibnamefont
  {Cacciari}}, \bibinfo {author} {\bibfnamefont {M.}~\bibnamefont {Greco}}, \
  and\ \bibinfo {author} {\bibfnamefont {P.}~\bibnamefont {Nason}},\ }\href
  {\doibase 10.1088/1126-6708/1998/05/007} {\bibfield  {journal} {\bibinfo
  {journal} {JHEP}\ }\textbf {\bibinfo {volume} {05}},\ \bibinfo {pages} {007}
  (\bibinfo {year} {1998})},\ \Eprint {http://arxiv.org/abs/hep-ph/9803400}
  {arXiv:hep-ph/9803400} \BibitemShut {NoStop}%
\bibitem [{\citenamefont {Cacciari}\ \emph {et~al.}(2001)\citenamefont
  {Cacciari}, \citenamefont {Frixione},\ and\ \citenamefont
  {Nason}}]{Cacciari:2001td}%
  \BibitemOpen
  \bibfield  {author} {\bibinfo {author} {\bibfnamefont {M.}~\bibnamefont
  {Cacciari}}, \bibinfo {author} {\bibfnamefont {S.}~\bibnamefont {Frixione}},
  \ and\ \bibinfo {author} {\bibfnamefont {P.}~\bibnamefont {Nason}},\ }\href
  {\doibase 10.1088/1126-6708/2001/03/006} {\bibfield  {journal} {\bibinfo
  {journal} {JHEP}\ }\textbf {\bibinfo {volume} {03}},\ \bibinfo {pages} {006}
  (\bibinfo {year} {2001})},\ \Eprint {http://arxiv.org/abs/hep-ph/0102134}
  {arXiv:hep-ph/0102134} \BibitemShut {NoStop}%
\bibitem [{\citenamefont {Eskola}\ \emph {et~al.}(2017)\citenamefont {Eskola},
  \citenamefont {Paakkinen}, \citenamefont {Paukkunen},\ and\ \citenamefont
  {Salgado}}]{Eskola:2016oht}%
  \BibitemOpen
  \bibfield  {author} {\bibinfo {author} {\bibfnamefont {K.~J.}\ \bibnamefont
  {Eskola}}, \bibinfo {author} {\bibfnamefont {P.}~\bibnamefont {Paakkinen}},
  \bibinfo {author} {\bibfnamefont {H.}~\bibnamefont {Paukkunen}}, \ and\
  \bibinfo {author} {\bibfnamefont {C.~A.}\ \bibnamefont {Salgado}},\ }\href
  {\doibase 10.1140/epjc/s10052-017-4725-9} {\bibfield  {journal} {\bibinfo
  {journal} {Eur. Phys. J. C}\ }\textbf {\bibinfo {volume} {77}},\ \bibinfo
  {pages} {163} (\bibinfo {year} {2017})},\ \Eprint
  {http://arxiv.org/abs/1612.05741} {arXiv:1612.05741 [hep-ph]} \BibitemShut
  {NoStop}%
\bibitem [{\citenamefont {Cao}\ \emph {et~al.}(2013)\citenamefont {Cao},
  \citenamefont {Qin},\ and\ \citenamefont {Bass}}]{Cao:2013ita}%
  \BibitemOpen
  \bibfield  {author} {\bibinfo {author} {\bibfnamefont {S.}~\bibnamefont
  {Cao}}, \bibinfo {author} {\bibfnamefont {G.-Y.}\ \bibnamefont {Qin}}, \ and\
  \bibinfo {author} {\bibfnamefont {S.~A.}\ \bibnamefont {Bass}},\ }\href
  {\doibase 10.1103/PhysRevC.88.044907} {\bibfield  {journal} {\bibinfo
  {journal} {Phys. Rev.}\ }\textbf {\bibinfo {volume} {C88}},\ \bibinfo {pages}
  {044907} (\bibinfo {year} {2013})},\ \Eprint {http://arxiv.org/abs/1308.0617}
  {arXiv:1308.0617 [nucl-th]} \BibitemShut {NoStop}%
%%CITATION = ARXIV:1308.0617;%%
\bibitem [{\citenamefont {Lin}\ \emph {et~al.}(2001)\citenamefont {Lin},
  \citenamefont {Di},\ and\ \citenamefont {Ko}}]{Lin:2000jp}%
  \BibitemOpen
  \bibfield  {author} {\bibinfo {author} {\bibfnamefont {Z.-w.}\ \bibnamefont
  {Lin}}, \bibinfo {author} {\bibfnamefont {T.~G.}\ \bibnamefont {Di}}, \ and\
  \bibinfo {author} {\bibfnamefont {C.~M.}\ \bibnamefont {Ko}},\ }\href
  {\doibase 10.1016/S0375-9474(00)00611-4} {\bibfield  {journal} {\bibinfo
  {journal} {Nucl. Phys.}\ }\textbf {\bibinfo {volume} {A689}},\ \bibinfo
  {pages} {965} (\bibinfo {year} {2001})},\ \Eprint
  {http://arxiv.org/abs/nucl-th/0006086} {arXiv:nucl-th/0006086 [nucl-th]}
  \BibitemShut {NoStop}%
%%CITATION = NUCL-TH/0006086;%%
\bibitem [{\citenamefont {Bass}\ \emph {et~al.}(1998)\citenamefont {Bass} \emph
  {et~al.}}]{Bass:1998ca}%
  \BibitemOpen
  \bibfield  {author} {\bibinfo {author} {\bibfnamefont {S.~A.}\ \bibnamefont
  {Bass}} \emph {et~al.},\ }\href {\doibase 10.1016/S0146-6410(98)00058-1}
  {\bibfield  {journal} {\bibinfo  {journal} {Prog. Part. Nucl. Phys.}\
  }\textbf {\bibinfo {volume} {41}},\ \bibinfo {pages} {255} (\bibinfo {year}
  {1998})},\ \Eprint {http://arxiv.org/abs/nucl-th/9803035}
  {arXiv:nucl-th/9803035 [nucl-th]} \BibitemShut {NoStop}%
%%CITATION = NUCL-TH/9803035;%%
\bibitem [{\citenamefont {Bleicher}\ \emph {et~al.}(1999)\citenamefont
  {Bleicher} \emph {et~al.}}]{Bleicher:1999xi}%
  \BibitemOpen
  \bibfield  {author} {\bibinfo {author} {\bibfnamefont {M.}~\bibnamefont
  {Bleicher}} \emph {et~al.},\ }\href {\doibase 10.1088/0954-3899/25/9/308}
  {\bibfield  {journal} {\bibinfo  {journal} {J. Phys.}\ }\textbf {\bibinfo
  {volume} {G25}},\ \bibinfo {pages} {1859} (\bibinfo {year} {1999})},\ \Eprint
  {http://arxiv.org/abs/hep-ph/9909407} {arXiv:hep-ph/9909407 [hep-ph]}
  \BibitemShut {NoStop}%
%%CITATION = HEP-PH/9909407;%%
\bibitem [{\citenamefont {Back}\ \emph {et~al.}(2003)\citenamefont {Back} \emph
  {et~al.}}]{Back:2002wb}%
  \BibitemOpen
  \bibfield  {author} {\bibinfo {author} {\bibfnamefont {B.}~\bibnamefont
  {Back}} \emph {et~al.},\ }\href {\doibase 10.1103/PhysRevLett.91.052303}
  {\bibfield  {journal} {\bibinfo  {journal} {Phys. Rev. Lett.}\ }\textbf
  {\bibinfo {volume} {91}},\ \bibinfo {pages} {052303} (\bibinfo {year}
  {2003})},\ \Eprint {http://arxiv.org/abs/nucl-ex/0210015}
  {arXiv:nucl-ex/0210015} \BibitemShut {NoStop}%
\bibitem [{\citenamefont {Adler}\ \emph {et~al.}(2004)\citenamefont {Adler}
  \emph {et~al.}}]{Adler:2003cb}%
  \BibitemOpen
  \bibfield  {author} {\bibinfo {author} {\bibfnamefont {S.}~\bibnamefont
  {Adler}} \emph {et~al.} (\bibinfo {collaboration} {PHENIX}),\ }\href
  {\doibase 10.1103/PhysRevC.69.034909} {\bibfield  {journal} {\bibinfo
  {journal} {Phys. Rev. C}\ }\textbf {\bibinfo {volume} {69}},\ \bibinfo
  {pages} {034909} (\bibinfo {year} {2004})},\ \Eprint
  {http://arxiv.org/abs/nucl-ex/0307022} {arXiv:nucl-ex/0307022} \BibitemShut
  {NoStop}%
\bibitem [{\citenamefont {Abelev}\ \emph {et~al.}(2008)\citenamefont {Abelev}
  \emph {et~al.}}]{Abelev:2008ae}%
  \BibitemOpen
  \bibfield  {author} {\bibinfo {author} {\bibfnamefont {B.~I.}\ \bibnamefont
  {Abelev}} \emph {et~al.} (\bibinfo {collaboration} {STAR}),\ }\href {\doibase
  10.1103/PhysRevC.77.054901} {\bibfield  {journal} {\bibinfo  {journal} {Phys.
  Rev. C}\ }\textbf {\bibinfo {volume} {77}},\ \bibinfo {pages} {054901}
  (\bibinfo {year} {2008})},\ \Eprint {http://arxiv.org/abs/0801.3466}
  {arXiv:0801.3466 [nucl-ex]} \BibitemShut {NoStop}%
\bibitem [{\citenamefont {Adamczyk}\ \emph {et~al.}(2012)\citenamefont
  {Adamczyk} \emph {et~al.}}]{Adamczyk:2012af}%
  \BibitemOpen
  \bibfield  {author} {\bibinfo {author} {\bibfnamefont {L.}~\bibnamefont
  {Adamczyk}} \emph {et~al.} (\bibinfo {collaboration} {STAR}),\ }\href
  {\doibase 10.1103/PhysRevD.86.072013} {\bibfield  {journal} {\bibinfo
  {journal} {Phys. Rev. D}\ }\textbf {\bibinfo {volume} {86}},\ \bibinfo
  {pages} {072013} (\bibinfo {year} {2012})},\ \Eprint
  {http://arxiv.org/abs/1204.4244} {arXiv:1204.4244 [nucl-ex]} \BibitemShut
  {NoStop}%
\bibitem [{\citenamefont {van Hecke}\ \emph {et~al.}(1998)\citenamefont {van
  Hecke}, \citenamefont {Sorge},\ and\ \citenamefont {Xu}}]{vanHecke:1998yu}%
  \BibitemOpen
  \bibfield  {author} {\bibinfo {author} {\bibfnamefont {H.}~\bibnamefont {van
  Hecke}}, \bibinfo {author} {\bibfnamefont {H.}~\bibnamefont {Sorge}}, \ and\
  \bibinfo {author} {\bibfnamefont {N.}~\bibnamefont {Xu}},\ }\href {\doibase
  10.1103/PhysRevLett.81.5764} {\bibfield  {journal} {\bibinfo  {journal}
  {Phys. Rev. Lett.}\ }\textbf {\bibinfo {volume} {81}},\ \bibinfo {pages}
  {5764} (\bibinfo {year} {1998})},\ \Eprint
  {http://arxiv.org/abs/nucl-th/9804035} {arXiv:nucl-th/9804035} \BibitemShut
  {NoStop}%
\bibitem [{\citenamefont {Zhou}(2017)}]{Zhou:2017ikn}%
  \BibitemOpen
  \bibfield  {author} {\bibinfo {author} {\bibfnamefont {L.}~\bibnamefont
  {Zhou}} (\bibinfo {collaboration} {STAR}),\ }\href {\doibase
  10.1016/j.nuclphysa.2017.05.114} {\bibfield  {journal} {\bibinfo  {journal}
  {Nucl. Phys. A}\ }\textbf {\bibinfo {volume} {967}},\ \bibinfo {pages} {620}
  (\bibinfo {year} {2017})},\ \Eprint {http://arxiv.org/abs/1704.04364}
  {arXiv:1704.04364 [nucl-ex]} \BibitemShut {NoStop}%
\bibitem [{\citenamefont {Adam}\ \emph {et~al.}(2020)\citenamefont {Adam} \emph
  {et~al.}}]{Adam:2019hpq}%
  \BibitemOpen
  \bibfield  {author} {\bibinfo {author} {\bibfnamefont {J.}~\bibnamefont
  {Adam}} \emph {et~al.} (\bibinfo {collaboration} {STAR}),\ }\href {\doibase
  10.1103/PhysRevLett.124.172301} {\bibfield  {journal} {\bibinfo  {journal}
  {Phys. Rev. Lett.}\ }\textbf {\bibinfo {volume} {124}},\ \bibinfo {pages}
  {172301} (\bibinfo {year} {2020})},\ \Eprint
  {http://arxiv.org/abs/1910.14628} {arXiv:1910.14628 [nucl-ex]} \BibitemShut
  {NoStop}%
\bibitem [{\citenamefont {Acharya}\ \emph {et~al.}(2018)\citenamefont {Acharya}
  \emph {et~al.}}]{Acharya:2018hre}%
  \BibitemOpen
  \bibfield  {author} {\bibinfo {author} {\bibfnamefont {S.}~\bibnamefont
  {Acharya}} \emph {et~al.} (\bibinfo {collaboration} {ALICE}),\ }\href
  {\doibase 10.1007/JHEP10(2018)174} {\bibfield  {journal} {\bibinfo  {journal}
  {JHEP}\ }\textbf {\bibinfo {volume} {10}},\ \bibinfo {pages} {174} (\bibinfo
  {year} {2018})},\ \Eprint {http://arxiv.org/abs/1804.09083} {arXiv:1804.09083
  [nucl-ex]} \BibitemShut {NoStop}%
\bibitem [{\citenamefont {Acharya}\ \emph {et~al.}(2019)\citenamefont {Acharya}
  \emph {et~al.}}]{Acharya:2018ckj}%
  \BibitemOpen
  \bibfield  {author} {\bibinfo {author} {\bibfnamefont {S.}~\bibnamefont
  {Acharya}} \emph {et~al.} (\bibinfo {collaboration} {ALICE}),\ }\href
  {\doibase 10.1016/j.physletb.2019.04.046} {\bibfield  {journal} {\bibinfo
  {journal} {Phys. Lett. B}\ }\textbf {\bibinfo {volume} {793}},\ \bibinfo
  {pages} {212} (\bibinfo {year} {2019})},\ \Eprint
  {http://arxiv.org/abs/1809.10922} {arXiv:1809.10922 [nucl-ex]} \BibitemShut
  {NoStop}%
\bibitem [{\citenamefont {Greco}\ \emph {et~al.}(2003)\citenamefont {Greco},
  \citenamefont {Ko},\ and\ \citenamefont {Levai}}]{Greco:2003mm}%
  \BibitemOpen
  \bibfield  {author} {\bibinfo {author} {\bibfnamefont {V.}~\bibnamefont
  {Greco}}, \bibinfo {author} {\bibfnamefont {C.~M.}\ \bibnamefont {Ko}}, \
  and\ \bibinfo {author} {\bibfnamefont {P.}~\bibnamefont {Levai}},\ }\href
  {\doibase 10.1103/PhysRevC.68.034904} {\bibfield  {journal} {\bibinfo
  {journal} {Phys. Rev. C}\ }\textbf {\bibinfo {volume} {68}},\ \bibinfo
  {pages} {034904} (\bibinfo {year} {2003})},\ \Eprint
  {http://arxiv.org/abs/nucl-th/0305024} {arXiv:nucl-th/0305024} \BibitemShut
  {NoStop}%
\bibitem [{\citenamefont {Cao}\ \emph {et~al.}(2020)\citenamefont {Cao},
  \citenamefont {Sun}, \citenamefont {Li}, \citenamefont {Liu}, \citenamefont
  {Xing}, \citenamefont {Qin},\ and\ \citenamefont {Ko}}]{Cao:2019iqs}%
  \BibitemOpen
  \bibfield  {author} {\bibinfo {author} {\bibfnamefont {S.}~\bibnamefont
  {Cao}}, \bibinfo {author} {\bibfnamefont {K.-J.}\ \bibnamefont {Sun}},
  \bibinfo {author} {\bibfnamefont {S.-Q.}\ \bibnamefont {Li}}, \bibinfo
  {author} {\bibfnamefont {S.~Y.~F.}\ \bibnamefont {Liu}}, \bibinfo {author}
  {\bibfnamefont {W.-J.}\ \bibnamefont {Xing}}, \bibinfo {author}
  {\bibfnamefont {G.-Y.}\ \bibnamefont {Qin}}, \ and\ \bibinfo {author}
  {\bibfnamefont {C.~M.}\ \bibnamefont {Ko}},\ }\href {\doibase
  10.1016/j.physletb.2020.135561} {\bibfield  {journal} {\bibinfo  {journal}
  {Phys. Lett. B}\ }\textbf {\bibinfo {volume} {807}},\ \bibinfo {pages}
  {135561} (\bibinfo {year} {2020})},\ \Eprint
  {http://arxiv.org/abs/1911.00456} {arXiv:1911.00456 [nucl-th]} \BibitemShut
  {NoStop}%
\bibitem [{\citenamefont {He}\ and\ \citenamefont {Rapp}(2020)}]{He:2019vgs}%
  \BibitemOpen
  \bibfield  {author} {\bibinfo {author} {\bibfnamefont {M.}~\bibnamefont
  {He}}\ and\ \bibinfo {author} {\bibfnamefont {R.}~\bibnamefont {Rapp}},\
  }\href {\doibase 10.1103/PhysRevLett.124.042301} {\bibfield  {journal}
  {\bibinfo  {journal} {Phys. Rev. Lett.}\ }\textbf {\bibinfo {volume} {124}},\
  \bibinfo {pages} {042301} (\bibinfo {year} {2020})},\ \Eprint
  {http://arxiv.org/abs/1905.09216} {arXiv:1905.09216 [nucl-th]} \BibitemShut
  {NoStop}%
\bibitem [{\citenamefont {Wang}\ \emph {et~al.}(2020)\citenamefont {Wang},
  \citenamefont {Song}, \citenamefont {Shao},\ and\ \citenamefont
  {Liang}}]{Wang:2019fcg}%
  \BibitemOpen
  \bibfield  {author} {\bibinfo {author} {\bibfnamefont {R.-Q.}\ \bibnamefont
  {Wang}}, \bibinfo {author} {\bibfnamefont {J.}~\bibnamefont {Song}}, \bibinfo
  {author} {\bibfnamefont {F.-L.}\ \bibnamefont {Shao}}, \ and\ \bibinfo
  {author} {\bibfnamefont {Z.-T.}\ \bibnamefont {Liang}},\ }\href {\doibase
  10.1103/PhysRevC.101.054903} {\bibfield  {journal} {\bibinfo  {journal}
  {Phys. Rev. C}\ }\textbf {\bibinfo {volume} {101}},\ \bibinfo {pages}
  {054903} (\bibinfo {year} {2020})},\ \Eprint
  {http://arxiv.org/abs/1911.00823} {arXiv:1911.00823 [hep-ph]} \BibitemShut
  {NoStop}%
\end{thebibliography}%

%%%% Slides
%%%% https://www.bnl.gov/aum2017/content/workshops/Workshop_1c/Rapp_OpenHeavyFlavorTheory.pdf
%%%% http://icts.ustc.edu.cn/chinese/seminar/transparencies/Guang-You%20Qin/Qin_jet_KeDa20160324.pdf

\end{document}